\documentclass[fleqn,usenatbib]{mnras}

\usepackage{newtxtext,newtxmath}

\usepackage[T1]{fontenc}

\DeclareRobustCommand{\VAN}[3]{#2}
\let\VANthebibliography\thebibliography
\def\thebibliography{\DeclareRobustCommand{\VAN}[3]{##3}\VANthebibliography}

\usepackage{graphicx}	% Including figure files
\usepackage{amsmath}	% Advanced maths commands

\usepackage{ulem}
\usepackage{xcolor}
\usepackage{hyperref}

\usepackage{subcaption}
\usepackage{booktabs}

\title[Regressing physical merger properties]
{Beyond classification: regressing physical merger properties using solid harmonic wavelet bispectrum}

\author[Brown et al.]{
Alex Brown$^{1}$, %\thanks{E-mail: ab3308@bath.ac.uk}
Georgios Exarchakis$^{1}$\thanks{E-mail: ge394@bath.ac.uk}, 
Mathilda Avirett-Mackenzie$^{2}$, 
Carolin Villforth$^{2}$\thanks{E-mail: c.villforth@bath.ac.uk}
\\
$^{1}$Department of Computer Science, University of Bath, BA2 7AY, Bath, UK\\   
$^{2}$Department of Physics, University of Bath, BA2 7AY, Bath, UK\\
}

\date{Accepted 6th November 2026. Received YYY; in original form ZZZ}

\pubyear{\the\year{}}

\begin{document}
\label{firstpage}
\pagerange{\pageref{firstpage}--\pageref{lastpage}}
\maketitle

\begin{abstract}
%intro
Galaxy mergers play an important role in galaxy evolution; they enhance star formation, transform morphologies, and may drive accretion onto black holes. Due to their complex morphological features, identification of mergers poses a significant challenge. Convolutional neural networks (CNNs) have been successful at identifying mergers with good accuracy. However, similar to other deep learning techniques such as Vision Transformers, they require large training samples. Current methods are also limited to classification, rather than regression. 
%method
Here, we apply the solid harmonic wavelet scattering bispectrum,
a method that encodes invariances and symmetries in the data which need to be learned by CNNs, as an alternative to CNNs.
%results
We apply our methodology to a simulated dataset of galaxies with a wide range of merger properties. We show that this method outperforms standard methodologies, while requiring fewer training samples and being robust to noise. Most importantly, we show that for the first time this method is able to regress on both merger mass ratio and time since merger simultaneously, with uncertainties $\sim 0.33$ dex uncertainty in merger mass ratio and $\sim 0.66$ Gyr uncertainty in time since merger.
%conclusions
We have therefore shown that solid harmonic wavelet bispectrum presents an alternative to currently used machine learning techniques.
\end{abstract}

\begin{keywords}
software: machine learning -- galaxies: structure -- galaxies: interactions -- galaxies: irregular
\end{keywords}

%%%%%%%%%%%%%%%%%%%%%%%%%%%%%%%%%%%%%%%%%%%%%%%%%%

%%%%%%%%%%%%%%%%% BODY OF PAPER %%%%%%%%%%%%%%%%%%

\section{Introduction}
\label{sec:intro}

Morphological analysis of images plays a major role in astrophysics. Galaxy morphologies can be used to determine the contribution of disk and bulges, but have also most commonly been used in the detection of galaxy mergers. Galaxy mergers cause strong disruptions of the distributions of gas and stars, resolution in features such a tidal bridges, double nuclei, tidal tails or shells \citep[e.g.][]{lotz_effect_2010, barnes_hernquist_1991}. The strength and type of merger feature changes significantly with the mass ratio and well as stage of merger \citep[e.g.][]{lotz_effect_2010, lotz_effect_2010-1, Scudder2015}.

Mergers play a major role in galaxy evolution, they disrupt galaxies, cause transformation from disk to bulge dominated system, cause strong starbursts, grow stellar as well as black hole mass, and may be involved in short phases of extreme growth of both the galaxy and the central black hole, after seminal work introducing the concept into galaxy evolution literature \citep{dimatteo_2005,hopkins_2008}, active research has refined our understanding of both the theoretical side \citep[e.g.][]{Renaud2022, Byrne2025, Mcalpine2020, Scudder2015} and observational work \citep[e.g.][]{pearson2025, Li2023}. Theoretically, research has shown the the impact of mergers is dependent on parameters of the galaxy itself, with specifically gas mass having a strong impact on start formation and black hole accretion \citep{lotz_effect_2010-1,lotz_effect_2010,cox2008,Byrne2025,Scudder2015}. On the other hand, observational work, including well constrained experiment comparing mergers to carefully matched control samples, have shown that the enhancement of star formation and black hole accretion evolves significantly during mergers \citep[e.g.][]{Satyapal2014,ellison_galaxy_2011}.  Additionally, comparisons of mergers to control samples has been able to disentangle the impact of mergers from those of galaxy properties such as central gas density and environment, showing the complexity of the role of mergers \citep{sabater2013,sabater2015}. There has also been increasing research into the impact of mergers of different mass ratios on  the overall galaxy population, given their relative rarity \citep[e.g.][and references therein]{Mcalpine2020,CV2023,martin2017,kaviraj_mergers}.

Given their complex impact on galaxy evolution, detecting mergers has therefore been of significant interest. However, due to the complex features seen in merger as well as the rarity of mergers, most simple modelling techniques that can be used to decompose galaxies are not suitable. Initial merger detection relied strongly on visual classification \citep[e.g.][]{lintott_galaxy_2008}. Human classifiers are able to spot and classify complex merger features, but with rising sample sizes, quantitative measures to detect mergers based on e.g. asymmetry, central concentration, smoothness, uneven flux distribution were developed and shown to be sensitive for different merger stages \citep[e.g.][]{conselice_fundamental_2006, pawlik_shape_2016,lotz_2004}. While quantitative measures are more easily comparable across samples and scale more easily to large sample sizes, they lack the ability of human classification to identify specific features. Human classification also requires significant person hours, but can now be upscaled to very large sample through the use of citizen scientists \citep{lintott_galaxy_2008,galaxyzoo2_2013}.

Machine learning has recently been introduced as a method combining the strength of quantitative measures and human classification \citep[e.g.][]{avirett-mackenzie_post-merger_2024,bickley_convolutional_2021,bottrell_deep_2019,ciprijanovic_deepmerge_2021,cheng_optimizing_2020,de_graaff_classifying_2025, dominguez-sanchez2018,dominguez-sanchez2019,euclidmorph2025,ferreira2024,huertas-company_contrastive2023,huertas-company_galaxy_2024,margalef-bentabol_galaxy_2024,pearson2019,walmsley_galaxy_2020,pearson2024,pearson2025, ferreira2026}. Machine learning scales to very large samples, while being able to detect complex features by training on human classification. Machine learning methods can also be trained on simulated galaxies \citep[e.g.][]{bottrell_deep_2019,avirett-mackenzie_post-merger_2024,pearson2019}, meaning rather than relying on potentially biased human classifiers, models can be trained on correctly labelled images. This has led to quick rise in use of machine learning in galaxy morphological analysis, including the analysis of mergers.

The most commonly used machine learning method for galaxy morphology has been Convolutional Neural Networks (CNNs) \citep[e.g.][]{huertas-company_galaxy_2024, bickley_convolutional_2021, margalef-bentabol_galaxy_2024}. CNNs learn hierarchical features through cascaded convolutional layers. They structurally encode translation invariance, but other physically relevant symmetries -- most importantly rotation invariance -- must be learned from data through augmentation. Vision Transformers and foundation models such as Zoobot \citep{walmsley_zoobot2023} have also been applied to galaxy morphology \citep[e.g.][]{euclidmorph2025, schechter_neurips_2025}, achieving strong performance but requiring substantially larger training sets due to their increased parameter counts.  Fine-tuning foundation models, for specific tasks, may reduce the data requirements but results in an operator that is more opaque as it is trained on different datasets with a different objective. Geometric deep learning approaches, such as Harmonic Networks \citep{harmonicnetworks_worral_2017}, partially address this by embedding equivariance to rotations directly in the architecture, reducing the data needed to learn such symmetries --- but still rely on learned filters fitted through stochastic optimisation, meaning that different training runs can yield different solutions.

A common limitation across all these approaches is the reduction of the physical problem to a classification task. As discussed earlier, the impact of mergers on the galaxy is complex and dependent on merger parameters. Until recently, most models are trained to assign discrete labels (merger/non-merger, or a small number of merger stages). Recently, several studies have attempted finer-grained classification into multiple merger stages \citep{margalef-bentabol_galaxy_2024, ferreira2024, de_graaff_classifying_2025}, but performance degrades significantly compared to binary classification \citep{margalef-bentabol_galaxy_2024}. \citep{ferreira2026} further increased the number of time bins when classifying mergers, showing that a more detailed analysis of merger properties is feasible, showing they can classify post-mergers into four time bins achieving accuracy exceeding 70\%. The model used combines traditional CNNs with Vision Transformers. \citet{pearson2024} performed regression on time since merger for a sample of major mergers, including also pre-mergers. They showed that CNNs are the most successful of their tested methods in recovering merger timescales, outperforming other methods. They showed that for the major merger sample, merger stage can be retrieved reliable, although late and early stages show larger uncertainty and asymmetric errors. The winning model is a CNN. more recent work shows that machine learning methods can be used to regress on merger features, helping to study more intricate effects of mergers that depend on merger properties.

A further practical constraint is data scarcity. The highest-quality training data comes from cosmological simulations, where merger properties are known exactly -- but simulation volumes are finite, limiting samples to $\sim 10^3$ mergers depending on redshift and mass ratio range \citep[e.g.][]{avirett-mackenzie_post-merger_2024, bickley_convolutional_2021,margalef-bentabol_galaxy_2024}, and the computational cost of running larger simulations is substantial. The alternative -- training on observed galaxies with human-assigned labels -- introduces its own limitations: visual classification is subjective, biased towards the most prominent merger features, and unreliable for low mass ratios and late merger stages \citep{avirett-mackenzie_post-merger_2024}. While data sizes and quality are set to increase with surveys such as e.g. Euclid or the Large Survey of Space and Time (LSST), those increases will be constrained by observational limitations. In either case, training data remains scarce relative to the requirements of modern deep learning architectures. Methods with fewer trainable parameters that can perform well on small datasets are therefore not merely desirable but necessary for progress in merger analysis. Beyond this practical consideration, there is a methodological argument for representations that encode known physical symmetries rather than learning them, since a model whose invariances are structurally guaranteed requires fewer assumptions to be validated numerically, and its predictions can be more directly connected to the physical properties it encodes.

In this paper, we aim to expand on the current literature by expanding regression to include both mass ratio and time since merger. We present a proof-of concept analysis, introducing solid harmonic wavelets as a machine learning technique with fewer parameters, a need for smaller training samples, and robustness to noise. The aim of this paper is to provide an introduction to this methodology and contrast it to current tools. Section \ref{sec:mergerdata} describes our dataset of simulated mergers and image realism. Section \ref{sec:method_rationale} offers an introduction to the rationale of our methodology, Sections \ref{sec:solid-harmonic-wavelet-scattering} - \ref{sec:scatteringbispectrum} describe our stand-alone wavelet-based architectures while section \ref{sec:deeplearning} describes deep learning benchmarks, along with hybrid architectures that utilise both wavelets and deep learning. In section \ref{sec:mergerclassificationresults}, we show the capabilities of wavelet-based approaches on the same binary labelled galaxy merger classification problem tackled by \cite{avirett-mackenzie_post-merger_2024} -- providing insight into motivations behind our methodologies before progressing to regression. Section \ref{sec:regressionidealisedresults} shows regression results on idealised data, while section \ref{sec:regressionnoisyresults} highlights the impact of realism on our proposed techniques. Further, we analyse correlations between $\mu$ and $\Delta\textrm{t}$ in section \ref{sec:correlation_mu_deltat}, providing meaningful connections between galaxy morphology and model performance. Finally, in section \ref{sec:featureimportance} we conduct a permutation importance analysis to provide exposition as to the most important morphological structures for regression on merger features when using our methodology.

\section{Simulated merger dataset}
\label{sec:mergerdata}

In this section, we discuss the creation of the simulated dataset used to train and test different methods for image analysis. Section \ref{sec:mergerselection} describes the selection of the merger sample from IllustrisTNG as well as derivation of merger mass ratios and time since merger used in the analysis. We then download idealised images to test algorithms on idealised data, we also create a suite of realistic data to test how the different methods perform on noisy data (see Section \ref{sec:images}). The focus of this study will be on high resolution high-SN data, as expected from e.g. Euclid or LSST.

\subsection{Selecting merger samples}
\label{sec:mergerselection}

To create a merger sample, we select mergers with a wide range of mass ratios and time since merger from IllustrisTNG, following methodology laid out in \cite{avirett-mackenzie_post-merger_2024}. We use merger trees from \citet{nelson_first_2018} and expand code from illustris-python\footnote{https://github.com/illustristng/illustris\_python} to select mergers from IllustrisTNG100 \citep{nelson_first_2018}. To minimise the impact of redshift of the analysis through evolution of galaxy properties, we select mergers for snapshot (95-99), corresponding to a redshift up to z=0.05. We select only completed mergers, rather than pre-mergers.

\begin{figure}
    \includegraphics[width=\columnwidth]{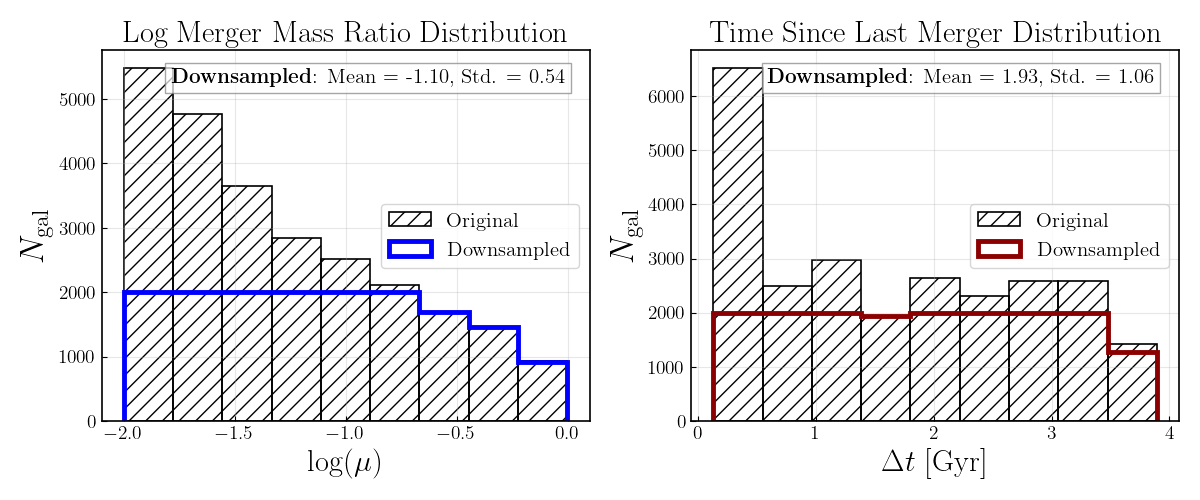}
    \centering
    \caption{Merger property distributions for merger mass ratio log($\mu$) (left) and time since merger $\Delta \textrm{t}$ in Gyr (right). Hashed histograms show the initial distributions, blue/red histograms the downsampled distributions used for training. Downsampling occurs after projections are separated into distinct training samples.}
    \label{fig:downsampled-target-distributions}
\end{figure}

The selected merger sample has a mass ratio ($\mu$) range of 1:1 to 1:100 and time since merger ($\Delta\textrm{t}$) up to 4 Gyr, distributions are shown in Figure \ref{fig:downsampled-target-distributions}. The limit on the mass ratio is chosen to include only mergers for which a significant impact on the morphology is expected. We choose the limit on $\Delta\textrm{t}$ irrespective on galaxy stellar mass.

The mass ratios $\mu$ and time since merger $\Delta\textrm{t}$ for the mergers are calculated as follows: Illustris TNG data is saved in snapshots, with time between snapshots $\sim$ 100--200 Myr over the range of snapshots used here. In individual snapshots, galaxies are identified using the SUBFIND algorithm \citep{springel2018} and traced through different snapshots using the SUBLINK algorithm \citep{rodriguez-gomez2015}. The uncertainty in the time since the merger $\Delta\textrm{t}$ is driven by the coarse time sampling of the simulation since the time of coalescence can be constrained only to occur between two snapshots, this yields an uncertainty of $\sim$ 100--200 Myr in this parameter. We calculate the stellar merger mass ratio following \citet{rodriguez-gomez2015}. We define the mass ratio $\mu$ as the ratio of the progenitor galaxy masses at the time prior to coalescence at which the smaller progenitor reaches its maximum stellar mass, thereby avoiding $\mu$ decreasing due to mass transfer.

Details on the independence between mass ration and time since merger can be found in Appendix \ref{app:ratio_time}.

\subsection{Idealised and realistic images}
\label{sec:images}

To evaluate merger image analysis, we use both idealised images, as well as image with realistic noise and sky background to evaluate the impact of noise on merger detection. We do not include full realistic backgrounds to our image analysis, but compare our methodology to existing methods to benchmark performance. A full analysis of the performance of our methodology is beyond the scope of the current paper, which focusses on the application of the new method. Following methodology from \citet{avirett-mackenzie_post-merger_2024}, we download idealised images from the 'Visualise galaxies and Halos' tool, see \citet{nelson_first_2018}, which projects stellar particles associated with the galaxy and its parent halo within the field of view (including any other galaxies within the same halo). The width of the image is set to 32 times the stellar half-mass radius. Stellar population synthesis models are used to predict the flux at a given wavelength \citep{conroy2009,conroy2010} and the fluxes are then convolved with filters to retrieve the magnitudes in different band. We use filters Hubble Space Telescope (HST) B/I/H (F438W, F814W, F160W) to cover a broad wavelength range. The images are initially created at the resolution of HST Advanced Camera for Surveys Wide Field Chanel (ACS WFC), which is 0.04 arcsec pix$^{-1}$, but are then rebinned to an image size of 128x128. Filters are all in rest frame, but due to the narrow redshift range, the impact of redshift on the observed wavelength is minimal. Please note that these visualizations do not include the effects of gas and dust in the ISM, however, gas and dust have been shown to not be necessary for reliable merger classification \citep{bottrell_deep_2019}. Each galaxy is imaged from three orthogonal viewing angles, giving three independent observations per galaxy.
We note that this does not include full radiative transfer and does not account for the effect of dust on colour. This is comparable to photometric semi-real dataset in \citep{bottrell_deep_2019} which shows good performance as a training set for full realistic data. We therefore do not expect this to significantly alter our results.

To test the performance of our methodology on realistic data (using code adapted from \citep{bottrell_deep_2019}\footnote{https://github.com/cbottrell/RealSim}), we simulate space based data with low noise levels and weak background, using a resolution of 0.1\arcsec, corresponding to $\approx$ 0.1kpc  arcsec$^{-2}$ for the redshift range analysed, slightly lower than the softening length of 0.7kpc at z $<1$ as well as sky levels of 25 mag arcsec$^{-2}$, for typical extragalactic near-infrared background. Poisson noise is then added representing a realistic range of of high-SN data, such as expected from e.g. Euclid or HST images. Below, we show how the level of noise affects the performance of the methodology.

\begin{figure}
    \centering
    \includegraphics[width=\columnwidth]{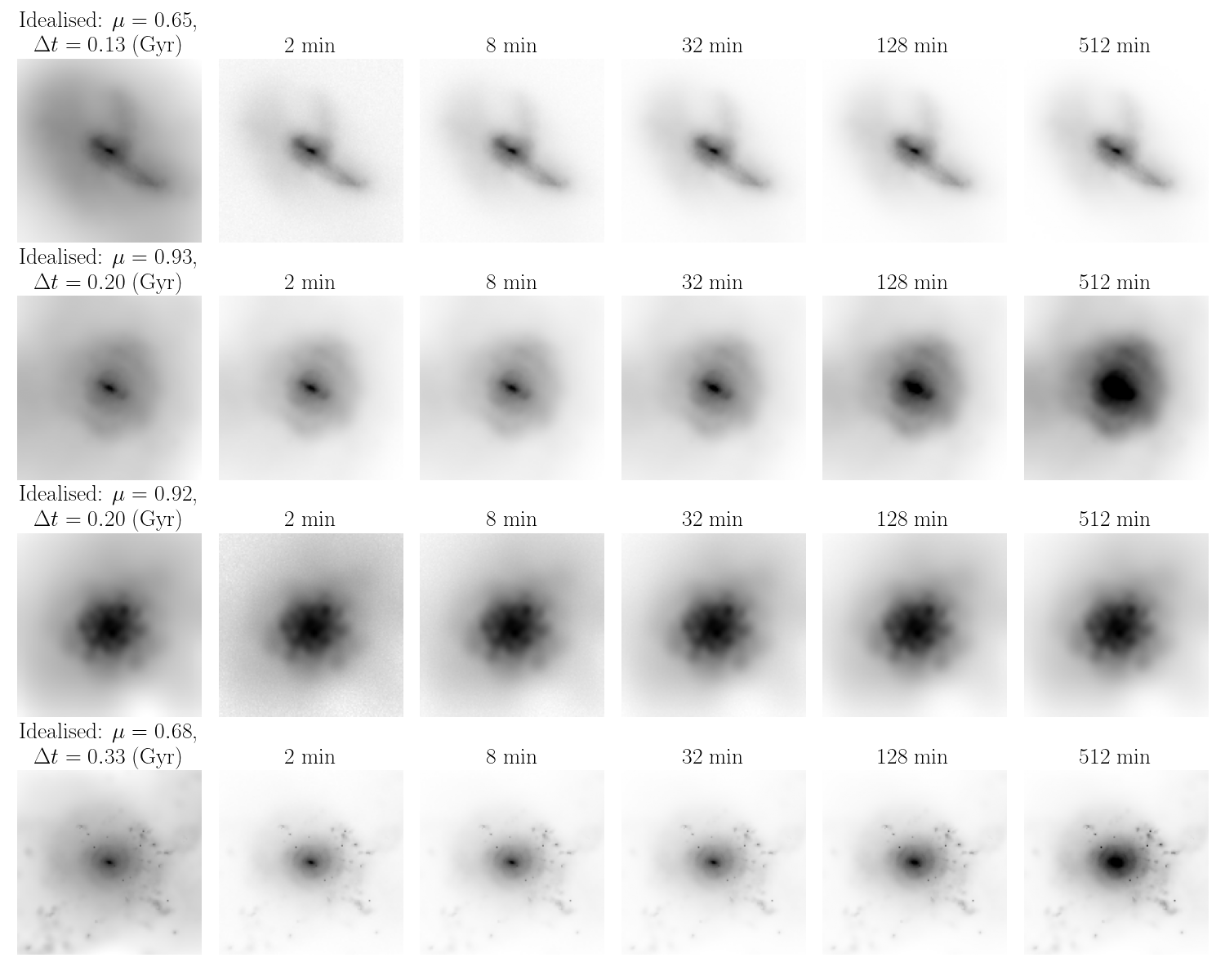}
   \caption{Example idealised and realistic galaxy images used throughout the paper. Leftmost image: idealised sample from IllustrisTNG simulation cropped to $32$ half stellar mass radius, and log normalised to $[0, 1]$. Left-to-right: the same sample with sky noise over increasing exposure times. Images are in AB magnitude.}
   \label{fig:merger-sample-increasing-exposure}
\end{figure}

\subsection{Dataset curation}
\label{sec:merger_train_test_description}

From the initial merger sample, we create two further task-specific datasets by undersampling majority histogram bins for both merger mass ratio and time since last merger coalescence. Making the data distribution more uniform equalises the variability of the data at all target ranges and provides more robust predictions using population based estimates.  We do so by creating a histogram of each target property, setting a maximum frequency count per bin and randomly selecting samples within each bin to conform to the specified size. We break both target properties (merger mass ratio and time since merger) into $16$ bins, each with a maximum of $2000$ samples, to create a balanced sample, while keeping the overall sample size large enough for the study. As in \cite{avirett-mackenzie_post-merger_2024}, for each galaxy, the three distinct viewing angles (projections) are used as individual samples. This results in full datasets consisting of $16,086$ and $17,216$ samples for $\mu$ and $\Delta\textrm{t}$, respectively. In our experiments, we reserve a $20\%$ held-out as a final test set. The remaining $80\%$ is used to calibrate and train a model through $k$-fold cross-validation. This approach allows us to ascertain how well models generalise across validation splits.

\section{Feature Extraction and Machine Learning Methods} \label{sec:featureextraction}

\subsection{Rationale}
\label{sec:method_rationale}

The following section provides an overview of the feature extraction methodology. Readers interested primarily in results may proceed to Section \ref{sec:results}. Full mathematical derivations and architectural details are given in Appendix \ref{apdx:background-knowledge}.

Deep Learning methods have excelled at mapping high-dimensional data to physically meaningful quantities by learning features from large training sets. Rather than learning features from data, we construct a fixed operator --- with no trainable parameters --- that encodes the relevant physical invariances by design. This is motivated by four requirements for regression on continuous merger properties:
 
\begin{enumerate}
    \item[(i)] \textbf{Rotation and translation invariance.} Viewing geometry carries no information about $\mu$ or $\Delta t$; the representation must be identical for all orientations of the same galaxy.
    \item[(ii)] \textbf{Multi-scale sensitivity.} Merger signatures span a wide range of physical scales, from compact double nuclei to extended tidal tails, all representing different physical aspects of mergers.
    \item[(iii)] \textbf{Cross-scale interactions.} The coupling between morphological structures at different scales --- for example, a disturbed outer envelope and a compact nucleus --- is more discriminative than either feature in isolation.
    \item[(iv)] \textbf{Deformation stability.} The representation must respond smoothly to the small morphological perturbations introduced by noise or modest changes in merger parameters.
\end{enumerate}
 
\noindent Solid harmonic wavelets \citep{eickenberg2017solid} transform predictably under rotation, yielding exact rotation invariance without data augmentation. Cascading them through a scattering network \citep{brunamallat2012scattering} provides multi-scale coverage and provable deformation stability. Taking the bispectrum of scattering responses --- the third order of spectral analysis --- captures quadratic interactions between morphological structures, while the angular frequency constraint causes rotational phase factors to cancel exactly, preserving invariance without discarding phase information. The resulting operator, the \textit{solid harmonic wavelet scattering bispectrum} (SHWSB), is entirely determined before any training data is seen. It is this property that enables strong performance in the data-scarce regime typical of merger studies (Section~\ref{sec:regressionidealisedresults}).
 
Full mathematical derivations are given in Appendix~\ref{apdx:background-knowledge}. In the remainder of this section, we summarise the key building blocks: solid harmonic wavelet scattering (Section~\ref{sec:solid-harmonic-wavelet-scattering}), its extension to the scattering bispectrum (Section~\ref{sec:scatteringbispectrum}), and the deep learning benchmarks against which we compare (Section~\ref{sec:deeplearning}).
 
The challenges of classification and regression also differ significantly in the complexity of the required representation. Binary merger classification requires learning a comparatively simple decision boundary, while regression on continuous properties demands modelling of subtle, continuous relationships between morphological features. In our framework, wavelet scattering already provides effective representations for classification (Section~\ref{sec:mergerclassificationresults}), while the scattering bispectrum captures higher-order interactions that yield the richer features needed for regression --- analogous to how deeper networks learn increasingly complex relationships.

\subsection{Solid Harmonic Wavelet Scattering} \label{sec:solid-harmonic-wavelet-scattering}
 
\begin{figure*}
    \centering
    \begin{subfigure}[T]{0.92\columnwidth}
        \centering
        \includegraphics[height=0.25\textheight]{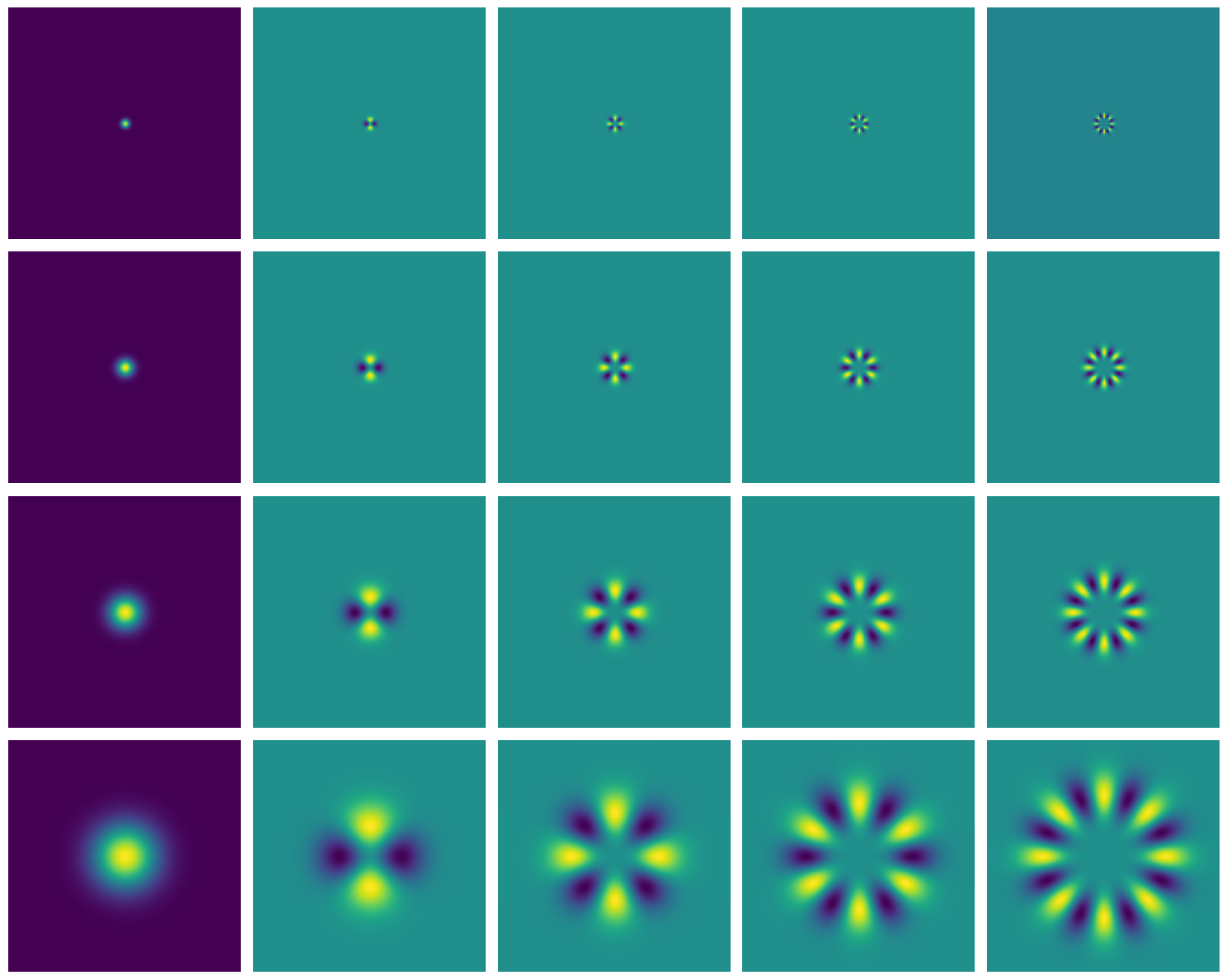}
        \caption{}
        \label{fig:filter-bank}
    \end{subfigure}
    \hspace{0.04\textwidth} 
    \begin{subfigure}[T]{0.92\columnwidth}
        \centering
        \includegraphics[height=0.25\textheight]{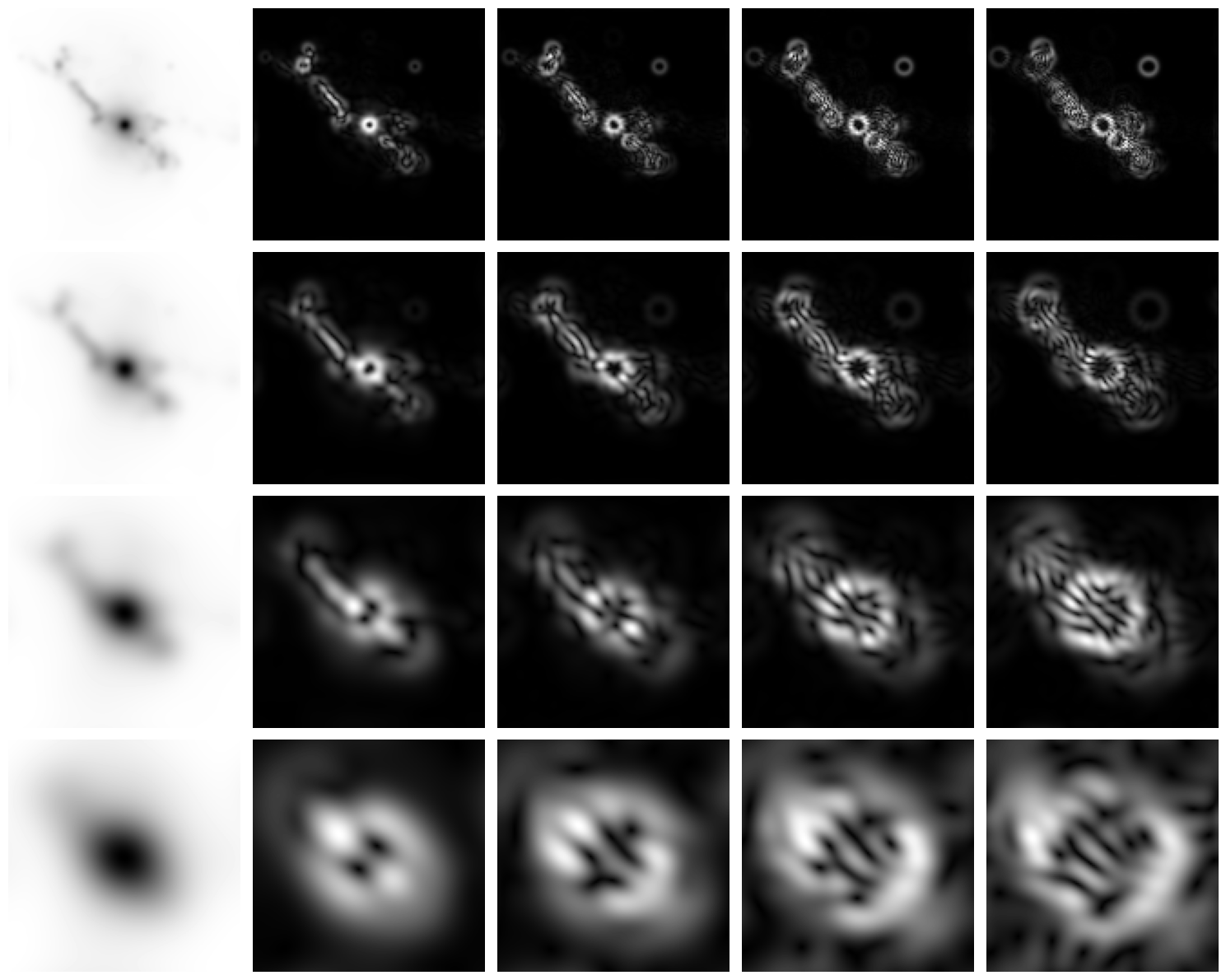}
        \caption{}
        \label{fig:features-extracted-wavelet-transform-l-j}
    \end{subfigure}
    \caption{
    Two-dimensional solid harmonic wavelet filters and their application to a merger. (a) Filter bank (real part). Left-to-right: angular frequency, $\ell$  increases from  (a Gaussian, $\ell=0$) in increments of $2$. Top-to-bottom: scale, $j$,  increases. (b) Wavelet transform responses for a noisy sample mocked to resemble the Euclid survey (1.8 hours exposure time), with filters corresponding to (a). Increasing scale results in broader, smoother features, while increasing angular frequency encodes finer directional structure. Images are in AB magnitude, log-normalised to $[0,1]$.}
    \label{fig:wavelet-filter-bank-and-result-of-wavelet-transform}
\end{figure*}

Solid harmonic wavelets \citep{eickenberg2017solid} form a complete, orthogonal basis on the unit circle, allowing galaxy images to be decomposed into rotational components without loss of information. In polar coordinates $(r, \varphi)$, they are parametrised by angular frequency $\ell$ and scale $\sigma$:
\begin{equation} \label{eq:solid-harmonic-wavelets}
    \psi_{\sigma,\ell}(r, \varphi) = \frac{1}{\sqrt{(2\pi\sigma^2)}^2} e^{-\frac{r^2}{2\sigma^2}} r^\ell e^{i \ell \varphi}
\end{equation}
We construct a filter bank over angular frequencies $\ell \in [0, L]$ and dyadic scales $\sigma = \sigma_{\text{mother}} \times 2^j$ for $j \in [0, J]$ (Figure~\ref{fig:filter-bank}). The maximum scale $2^J$ is chosen to match the largest features relevant to the task, ensuring coverage from fine local structure to extended global features. The wavelet transform of a signal $x$ is the convolution $W[x, j, \ell] = x \star \psi_{j, \ell}$, and solid harmonics guarantee that this transform is \textit{rotation-covariant}: rotating the input by angle $\gamma$ multiplies the response by a phase factor $e^{i\ell\gamma}$ while applying the same spatial rotation \citep{eickenberg2017solid}:
\begin{equation}
    W[R_\gamma x, j, \ell] = e^{i\ell\gamma} R_\gamma W[x, j, \ell] \label{eq:signal-rotation-commutes}
\end{equation}
Applying the complex modulus eliminates this phase factor, and summing over spatial coordinates then yields features that are fully invariant to both rotation and translation.
 
The \textit{Solid Harmonic Wavelet Scattering} (SHWS) transform cascades these wavelet transforms with modulus nonlinearities, analogously to Morlet scattering \citep{brunamallat2012scattering} but with built-in rotational structure. Zeroth-, first-, and second-order scattering coefficients capture progressively richer information:
\begin{align}
    & S_0[x,p] = \| x \|_p \label{eq:zeroth-order-shws} \\
    & S_1[x,j_1, \ell, p] = \| x \star \psi_{j_1, \ell} \|_p \label{eq:first-order-shws} \\
    & S_2[x,j_1, j_2, \ell, p] = \| \, | x \star \psi_{j_1, \ell} | \star \psi_{j_2, \ell} \, \|_p \label{eq:second-order-shws}
\end{align}
where $\|\cdot\|_p$ denotes $L_p$ pooling over spatial coordinates. Zeroth-order coefficients capture global signal energy, first-order coefficients encode spatial structure at each scale, and second-order coefficients recover cross-scale interactions lost through the modulus. The constraint $j_1 < j_2$ avoids redundancy: the modulus operator smooths the signal, attenuating high-frequency content, so that a second convolution at a scale smaller than the first would capture little additional information \citep{mallat2012groupinvariantscattering}.
 
Figure~\ref{fig:features-extracted-wavelet-transform-l-j} illustrates how varying $\ell$ and $j$ extracts complementary morphological information: increasing scale captures progressively larger structures with a smoothing effect, while increasing angular frequency encodes finer directional patterns. A comparison with oriented Morlet wavelets is provided in Appendix~\ref{apdx:oriented_wavelets}.

\subsection{Solid Harmonic Wavelet Scattering Bispectrum} \label{sec:scatteringbispectrum}

\subsubsection{Solid Harmonic Wavelet Bispectrum and Bicoherence}\label{sec:scatteringbicoherence}
The Solid Harmonic Wavelet Bispectrum (SHWB) \citep{ourpaper} is defined as the product of solid harmonic wavelet responses for $\ell_1, \ell_2$ and the complex conjugate of the wavelet response for $\ell_3 = \ell_1 + \ell_2$ at scale $j$: 
\begin{align}
    \text{SHWB}[x, j, \ell_1, \ell_2] & = W[x, j, \ell_1] \cdot W[x, j, \ell_2] \cdot \overline{W[x, j, \ell_{3}]} \label{solid harmonic wavelet bispectrum}
\end{align}
The Solid Harmonic Wavelet Bicoherence (SHWBic) is computed by normalising the bispectrum with respect to its magnitude:
\begin{align}
    & \text{SHWBic}[x, j, \ell_1, \ell_2] = \nonumber
    \\
    & \quad \frac{W[x, j, \ell_1] \cdot W[x, j, \ell_2] \cdot \overline{W[x, j, \ell_3]}}{\| W[x, j, \ell_1] \| \cdot \| W[x, j, \ell_2] \| \cdot \| \overline{W[x, j, \ell_3]} \| }
\end{align}
This normalisation removes the amplitude dependence, isolating the strength of phase alignment between components. As with the classical bispectrum, SHWBic takes values in $[0, 1]$, where values closer to 1 indicate strong phase coupling and structured nonlinear interactions. Since the bicoherence is a scalar normalisation of the bispectrum, it inherits the same covariance properties derived below. In our experiments, we use the bicoherence (SHWBic) rather than the bispectrum, as the normalisation makes features less sensitive to contrast variations and empirically yields better regression performance. Throughout this paper, we refer to the general methodology as the solid harmonic wavelet bispectrum, with the understanding that bicoherence is used in practice.

As shown in \cite{eickenberg2017solid}, a rotation $R_\gamma \in \mathrm{SO(2)}$ applied on signal $x$ of angle $\gamma$ produces the same rotation on its solid harmonic wavelet convolution, multiplied by a phase factor encoding the rotation angle \eqref{signal rotation commutes}. This arises due to the angular dependence of the wavelet on $\ell$.
\begin{align}
    W[R_\gamma x, j, \ell] & = e^{i\ell\gamma} R_\gamma W[x, j, \ell] \label{signal rotation commutes}
\end{align}

Substituting \eqref{signal rotation commutes} into \eqref{solid harmonic wavelet bispectrum}, and noting that the angular phase terms cancel due to the frequency constraint $\ell_3 = \ell_1 + \ell_2$, we obtain rotational covariance of SHWB:
\begin{align}
    \text{SHWB}[R_\gamma x, j, \ell_1, \ell_2] &= R_\gamma \text{SHWB}[x, j, \ell_1, \ell_2] \label{bispectrum covariance proof}
\end{align}

In contrast to scattering networks with Morlet wavelets, where rotational invariance is achieved by discarding phase information via a modulus, SHWB retains complex phase while still achieving covariance. Since convolutional operators are also translation-covariant \cite{mallat2012groupinvariantscattering}, we can obtain roto-translation invariance by applying spatial averaging or other reduction operations, such as $L_p$ pooling.

\subsubsection{Integration with wavelet scattering}

We now integrate the bispectrum with wavelet scattering, yielding the Solid Harmonic Wavelet Scattering Bispectrum (SHWSB) and its normalised variant, the Solid Harmonic Wavelet Scattering Bicoherence (SHWSBic) \citep{ourpaper}. While SHWS produces roto-translation invariant features and linearises deformations, it does so by discarding complex phase information which can carry discriminative structural information — which SHWSB retains. The bispectrum of solid harmonic wavelet scattering coefficients is therefore defined as the triple-product of intermediate scattering responses with respect to angular frequency. By design, it achieves the same roto-translation covariance properties without applying the modulus operator. This provides deformation stability through scattering and higher-order structure from the bispectrum. Following Figure \ref{fig:features-extracted-wavelet-transform-l-j}, whereby changing the angular frequency changes the features extracted, it can more abstractly be viewed as encoding the interactions between feature types --- or more precisely, morphological structures.

The zeroth-order bispectral coefficient is computed through $L_p$ pooling to capture global energy:
\begin{align}
    \text{SHWSB}_0[x,p] & = \| x \|_p \label{zeroth order scattering bispectrum}
\end{align}
We introduce the following  notation for the linear responses of the first- and second-order scattering:
\begin{align}
    & W_1[x,j_1, \ell] = x \star \psi_{j_1, \ell} \label{first order scattering bispectrum} \\
    & W_2[x,j_1, j_2, \ell] = | x \star \psi_{j_1, \ell} | \star \psi_{j_2, \ell} \label{second order scattering bispectrum}
\end{align}
Note that we no longer apply the modulus operator as in Equations \ref{eq:first-order-shws}--\ref{eq:second-order-shws}. For all $\ell_1, \ell_2 \in L$ with $\ell_1 < \ell_2$, and $\ell_1 + \ell_2 = \ell_3 \in L$, we define the bispectral coefficients:
\begin{align}
    & \text{SHWSB}_1[x, j_1, \ell_1, \ell_2] = \label{SHWSB1} \\
    & \quad W_1[x,j_1, \ell_1] \cdot W_1[x,j_1, \ell_2] \cdot \overline{W_1[x,j_1, \ell_{3}]} \nonumber \\
    & \text{SHWSB}_2[x, j_1, j_2, \ell_1, \ell_2] = \label{SHWSB2} \\
    & \quad W_2[x,j_1, j_2, \ell_1] \cdot W_2[x, j_1, j_2, \ell_2] \cdot \overline{W_2[x,j_1, j_2, \ell_{3}]} \nonumber
\end{align}
The bicoherence (SHWSBic) is derived by normalising with the product of the scattering magnitudes:
\begin{align}
    & \text{SHWSBic}_1[x, j_1, \ell_1, \ell_2] = \label{SHWSBic1} \\  
    & \quad \frac{W_1[x,j_1, \ell_1] \cdot W_1[x,j_1, \ell_2] \cdot \overline{W_1[x,j_1, \ell_3]}}{\| W_1[x, j_1, \ell_1] \| \cdot \| W_1[x, j_1, \ell_2] \| \cdot \| W_1[x, j_1, \ell_3] \|} \nonumber
\end{align}
\begin{align}
    & \text{SHWSBic}_2[x, j_1, j_2, \ell_1, \ell_2] = \label{SHWSBic2} \\  
    & \quad \frac{W_2[x,j_1, j_2, \ell_1] \cdot W_2[x, j_1, j_2, \ell_2] \cdot \overline{W_2[x,j_1, j_2, \ell_{3}]}}
    {\| W_2[x,j_1, j_2, \ell_1] \| \cdot \| W_2[x,j_1, j_2, \ell_2] \| \cdot \| \overline{W_2[x, j_1, j_2, \ell_{3}]} \|} \nonumber
\end{align}
As in Section \ref{sec:scatteringbicoherence}, this normalisation isolates phase alignment from amplitude, and SHWSBic inherits the same $[0,1]$ range and covariance properties. By factoring out local energy, SHWSBic is less sensitive to contrast variations and better suited for discriminative tasks. Thus, it quantifies the alignment between rotational symmetries in galaxy morphology. As with the non-scattering case, the bicoherence inherits the covariance properties of the bispectrum. In our regression experiments, we use SHWSBic as it empirically yields better performance.

We extract features from Equations \ref{SHWSB1}--\ref{SHWSBic2} by applying $L_p$ pooling on the spatial coordinates of the representations, which we denote with the parameter $p$ as:
\begin{align}
    \text{SHWSB}_1[x, j_1, \ell_1, \ell_2, p] & = \| \text{SHWSB}_1[x, j_1, \ell_1, \ell_2] \|_p 
\end{align}
By constraining the angular frequencies used in bispectral calculations, we observe that the phase factor arising from the solid harmonic wavelet transform cancels out (see Appendix \ref{apdx:shwb-covariance-proof}). Since convolutional operators are also translation-covariant \citep{mallat2012groupinvariantscattering}, it is clear that first-order coefficients are roto-translation covariant by definition. Meanwhile, intermediate second-order scattering coefficients in \eqref{second order scattering bispectrum} retain the modulus nonlinearity after the first convolution, preserving stability. This cancels the complex phase factor arising in \eqref{eq:signal-rotation-commutes}, leaving:
\begin{align}
U[j, \ell]R_\gamma x(u) = R_\gamma (x \star \psi_{j, \ell})(u) \label{rotation left after modulus - scattering bispectrum}
\end{align}
Since solid harmonics commute with rotation, a second convolution preserves rotation structure. Therefore, the earlier angular frequency constraint establishes that both first- and second- order SHWSB coefficients are roto-translation covariant. $L_p$ pooling reduces these to roto-translation invariant features.

The scattering bispectrum captures interactions between responses with respect to angular frequency, while the bicoherence quantifies the strength of such interactions. Since we have demonstrated in Figure \ref{fig:features-extracted-wavelet-transform-l-j} that varying angular frequency alters the features extracted, this can more abstractly be viewed as interactions between  morphological features --- for example, the coupling between tidal tails and the central bulge.

\subsection{Deep Learning} \label{sec:deeplearning}

We benchmark both euclidean and non-euclidean convolutional neural networks, namely CNNs and Harmonic Networks, in addition to hybrid architectures that leverage wavelet scattering as a robust initial feature extractor followed by learnable filters. This allows us to place the performance of our wavelet-based approaches against current methodologies applied in astrophysics (e.g. CNNs) and more data-efficient architectures (e.g. Harmonic Networks) applied in machine learning. We optimise architectures subject to dataset size and the relevant task (classification versus regression). Here, we provide high-level overviews of the architectures while detailed breakdowns are in Appendix \ref{apdx:deep-learning-architectures}. 

\subsubsection{Convolutional Neural Networks}
\label{sec:cnns}

Convolutional Neural Networks (CNNs) have achieved remarkable success in a wide range of tasks by learning hierarchical features through cascaded convolutional layers, nonlinear activations, and pooling operations. In the context of galaxy morphology, early convolutional layers capture low-level structural features such as tidal tails, spiral arms, and central bulges, while deeper layers progressively learn abstract morphological representations relevant to the chosen task. 

We reimplement the CNN architecture from \cite{avirett-mackenzie_post-merger_2024} for merger classification. The architecture is comprised of four convolutional blocks each containing convolutional layers with varying sized kernels, ReLU activation, batch normalisation for training stability and dropout for regularisation. Feature map channels increase with each successive convolutional block, to increase representational capacity at deeper layers, while the third block employs max pooling to downsample the feature map, offering some local invariance to transformations. The final convolutional features are flattened and processed through a two-layer multilayer perceptron with ReLU activation between layers to perform binary galaxy merger classification. For binary merger classification, we increase the count of high-mass galaxies using mirroring and rotations - as done by \cite{avirett-mackenzie_post-merger_2024}. We also apply the same train-time augmentations consisting of random horizontal and vertical flips, rotations of up to $15^{\circ}$, and small translations of up to a few pixels.

Further, for regression on merger features, it is reasonable to assume that an increasingly complex decision plane must be learnt. Thus, we increase the model capacity using further convolutional blocks and a deeper multilayer perceptron. This results in a roughly $8\times$ increase in learnable parameters compared to the CNN used by \cite{avirett-mackenzie_post-merger_2024}.

\subsubsection{Harmonic Networks}
\label{sec:h-nets}

Harmonic Networks (H-Nets) \citep{harmonicnetworks_worral_2017} embed equivariance by replacing filters with complex-valued, steerable circular harmonics. Circular harmonics transform predictably under rotation, resulting in complex phase shifts. This means that H-Nets can handle continuous rotations. Standard convolution operators in CNNs are replaced by cross-correlation. Since circular harmonics are complex-valued, each filter is decomposed into real and imaginary components which are treated as two separate channels. Multiple filters with different angular frequencies are used per layer, generating feature maps that respond to different rotational symmetries.

While H-Nets handle continuous rotations and allow for more efficient learning of discriminative features compared to CNNs, they increase the complexity of already convoluted architectures. It requires careful construction of the filters a priori, with the increased computational overheads of deep learning. Their filters -- and subsequent performance on downstream tasks -- are heavily influenced by both the number of rings and rotation order (angular frequency). 

Rotation order can be viewed analogously to the angular frequency parameter of two-dimensional solid harmonics, dictating the number of sinusoidal oscillations and thus capturing varying levels of rotational symmetry. Therefore, increasing the rotation order gives rise to a higher-capacity network capable of capturing increasingly complex features. Since H-Nets maintain separate feature maps for each order, this dramatically increases the number of channels flowing through the network and thus the number of learnable parameters. 

Meanwhile, filters are defined over concentric rings around the filter centre, where each ring corresponds to pixels at a fixed radius. The learnable radial profile weights each ring, controlling filter intensity. Thus, increasing the number of rings gives rise to a finer radial resolution. At the cost of increased computational complexity, this allows finer features to be extracted. 

In our experiments, we find that for increasingly complex decision boundaries, increasing both rotation order and number of rings is necessary, providing a model with capacity to learn a richer feature space. In such cases, we also increase the number of filters in each convolutional block. Comparatively, for smaller training set sizes, a much simpler architecture with low angular frequency is required to prevent overfitting. H-Nets serve as a good benchmark for equivariant deep learning architectures, which leverage known symmetries to more efficiently learn patterns in the data compared to conventional CNNs, and thus typically perform better in smaller data regimes.

\subsubsection{Hybrid architectures}
\label{sec:hybrid-architectures}

Hybrid architectures \citep{Oyallon_2019_hybrid, zarka2020deep_hybrid} that employ wavelet scattering as an initial feature extractor have demonstrated results comparable to, and sometimes exceeding, those of stand-alone deep learning networks such as CNNs, while requiring fewer trainable parameters. This motivates our use of wavelet-based feature extraction: scattering transforms provide desirable mathematical guarantees such as local translation-invariance and stability to small deformations. These properties are crucial to downstream tasks such as the analysis of astronomical images. Further, oriented Morlet wavelet scattering provides translation-invariance, while solid harmonic wavelet scattering provides rotation- and translation-covariance, with full invariance achieved through pooling.

Thus, wavelet scattering abstracts the need for a deep learning architecture to learn important symmetries such as rotation- and translation-covariance, while simplifying the decision boundary with respect to deformation. Therefore, we also benchmark a hybrid CNN in which Morlet wavelet scattering is used as an initial feature extractor. We parametrise this such that scattering feature maps serve as fixed low-level descriptors, followed by convolutional layers for higher-level feature extraction. We construct a Morlet wavelet filter bank up to scale $J=2$ sampled over $8$ angular orientations, which is used in a scattering network up to the second order. The scattering network produces an output of dimension $(B,C,1+LJ+\frac{L^2J(J-1)}{2},\frac{M}{2^J},\frac{N}{2^J})$ \citep{kymatio}, which we reshape to $(B,K,\frac{M}{2^J},\frac{N}{2^J})$ for compatibility with PyTorch convolutional layers. The scattering procedure downsamples features to $\frac{36}{2^J} \times \frac{36}{2^J} = 9 \times 9$ for each channel, when used for galaxy classification where training images are $36\times36$ over green, red and infrared channels. Features produced by the scattering transform are stable to deformation, locally translation-invariant and rotation-covariant. This means that in our subsequent learned convolutional filters, we are training on a simplified representation and, as such, can more efficiently learn key symmetries in the data.

We employ a pointwise convolution layer to reduce the channel dimension, followed by depthwise convolutional blocks with ReLU activation to refine the feature space. Batch normalisation and dropout help to stabilise the training procedure while reducing overfitting. Features are flattened before being classified by the same multilayer perceptron as used by \cite{avirett-mackenzie_post-merger_2024}. This offers a fair comparison of the architectures' ability to extract meaningful features without simply scaling up the classification unit. 

In short, wavelet scattering extracts generic features of varying granularities using predefined filters while learnable convolutional filters further refine these through back-propagation for the chosen task. In this work, we demonstrate this model's capabilities on merger classification.

\section{Results}
\label{sec:results}

Here, we present the application of the methodology introduced here, SHWBic, compared to other methods used in the field. Section \ref{sec:mergerclassificationresults} compares performances using a binary classification task, Section \ref{sec:regressionidealisedresults} shows the performance of our methodology on regression of merger features on idealised data, Section \ref{sec:regressionnoisyresults} shows the regression performance on noisy images and analyses the impact of noise on the performance. We discuss a correlation between time since merger and merger mass ratio detected in the regression in Section \ref{sec:correlation_mu_deltat}. Finally, we discuss interpretation of feature importance and the potential application to explainability in Section \ref{sec:featureimportance}.

\subsection{Merger Classification} \label{sec:mergerclassificationresults}

We benchmark wavelet-based techniques on the binary merger classification task defined by \cite{avirett-mackenzie_post-merger_2024}, using the same $80:10:10$ split ratio for training, validation, and test sets across all methods. Here, we use a sample of mergers and non-mergers, defined as in \citet{avirett-mackenzie_post-merger_2024}, where mergers are defined as having undergone a merger with mass ratio $\mu \geq 0.25$ in the last 500Myr and non-mergers are defined as having had no merger with mass ratio $\mu \geq 0.01$ in the last 2Gyr. Classes were balanced as in \citep{avirett-mackenzie_post-merger_2024}, resulting in $\approx 6000$ images in the merger and control sample. Any objects in between are not used for the binary classification. Details can be found in \citep{avirett-mackenzie_post-merger_2024}. Our methodology only varied in the following aspects:

\begin{itemize}
    \item An \textbf{increased initial learning rate} of $0.001$ (as opposed to $0.0001$), allowing faster model convergence with negligible performance degradation.
    \item A \textbf{learning rate scheduler} that reduces the learning rate after $7$ epochs without a decrease in the validation loss (\textit{Binary Cross-Entropy}), to guide towards more efficient convergence when the training plateaus.
    \item \textbf{Early stopping} after $25$ epochs (as opposed to $50$ epochs) without a decrease in the validation loss, reducing computational cost required for marginal accuracy gains towards the end of the training cycle.
\end{itemize}

\begin{table}
    \centering
    \footnotesize
    \setlength{\tabcolsep}{3pt}
    \caption{ Merger classification on the held-out test set ($n = 1379$; 681 mergers, 698 non-mergers) comparing wavelet-based techniques with a benchmark CNN, alongside the number of trainable parameters. Precision and recall are for the merger class. Results obtained on the same train-test splits.Train-time augmentations are only applied to the CNN, and hybrid Morlet Scattering + CNN. We report the mean $\pm$ standard deviation of the metrics over five training runs with different seeds.}
    \label{tab:classification-results}
    \begin{tabular}{lcccr}
    \toprule
        \textbf{Model} & \textbf{Acc. (\%)} & \textbf{Prec.} & \textbf{Rec.} & \textbf{Params} \\
    \midrule
        Linear (Morlet)      & $68.9 \pm 0.8$ & $0.68 \pm 0.01$ & $0.69 \pm 0.02$ & 2.9k \\
        Linear (SHWS)        & $74.6 \pm 0.1$ & $0.74 \pm 0.00$ & $0.74 \pm 0.00$ & 638 \\
    \midrule
        CNN                  & $78.8 \pm 0.8$ & $0.79 \pm 0.01$ & $0.78 \pm 0.01$ & 1.15M \\
        Hybrid (Morlet)      & $78.2 \pm 0.9$ & $0.79 \pm 0.02$ & $0.76 \pm 0.01$ & 172k \\
    \bottomrule
    \end{tabular}
\end{table}

Table \ref{tab:classification-results} shows that wavelet scattering features substantially linearise the classification problem. A single linear layer with $L_2$ regularisation on Morlet scattering features achieves $68.9\%$  accuracy, while solid harmonic wavelet scattering (SHWS) features improve this to $74.6\%$ --- with only $638$ trainable parameters and no data augmentation. The improvement reflects the benefit of built-in rotational invariance. Neither linear model matches the $78.8\%$ accuracy of the CNN from \cite{avirett-mackenzie_post-merger_2024}, but the gap is notably small given the difference in model complexity. The linear model is convex with a unique and analytically obtainable solution whereas the CNN requires nonconvex optimisation with no guarantees of reaching a global optimum.

Combining wavelet scattering with learned nonlinear layers closes this gap. A hybrid architecture using Morlet scattering as an initial feature extractor followed by convolutional layers achieves $78.2\%$
accuracy --- statistically comparable to the standalone CNN while using almost ten-fold fewer trainable parameters. Scattering features thus provide a strong foundation that reduces the burden on learned components.

These results confirm that solid harmonic wavelet scattering captures physically meaningful structure from merger images. Binary classification, however, provides limited insight into merger properties. The main result we wish to pursue is regression on merger mass ratio and time since merger, where the interpretable representations provided by solid harmonic wavelets become essential.

\subsection{Regression on idealised data} \label{sec:regressionidealisedresults}

Moving on from binary classification, we now explore regression on merger properties, specifically merger mass ratio $\mu$ and time since merger $\Delta\textrm{t}$, using the data described in Section \ref{sec:mergerdata}. Deriving the full physical properties of the merger (timescale and mass ratio), rather than a simple binary classification, would significantly improve the scientific impact of merger classification and understanding of the underlying physics.

We explore two regression approaches: linear regression with $L_2$ regularisation (fit using singular value decomposition), and k-nearest-neighbours (KNN) regression. The KNN regressor finds the $k$ training samples closest to a query point using $L_1$ (Manhattan) distance in feature space, then predicts the target value as a weighted average of the neighbours' targets, where each neighbour is weighted by the inverse of its distance — giving closer neighbours greater influence on the prediction.

Table \ref{tab:galaxy-regression-results} shows that models trained on our representations effectively regress merger features in idealised samples without observational noise or sky addition with a typical error of $\sim$ 0.3 dex over 2 orders of magnitude in merger mass ratio. We compare our approaches to a baseline random predictor, where predictions are sampled from the target range - demonstrating an almost 50\% reduction in error rates.

We observe an improvement using SHWS over oriented Morlet wavelets when input to linear models, reinforcing our belief that Solid Harmonic wavelets are better suited to domains in which we desire an invariance to uninformative transformations. Switching from linear regression to KNN regression on SHWS features yields further improvement, and the best performance is achieved using SHWSBic features with KNN regression. This suggests that accurate prediction of merger features requires capturing not just global symmetries but also complex relationships between features at varying scales — precisely what the SHWSBic encodes. This relationships between features on varying scales directly encode relations between merger features (e.g. asymmetries, double cores, tidal tails) on different physical scales. The non-parametric KNN approach captures additional nonlinearities from our coefficients without imposing restrictive functional forms. 

\begin{table}
    \small
    \centering
    \caption{Results of merger feature regression on $128\times128$ idealised samples. Images on a log scale over AB magnitude.}
    \label{tab:galaxy-regression-results}
    \begin{tabular}{lcccc}
    \toprule
        \textbf{Model} & \multicolumn{2}{c}{\textbf{Log $\mu$}} & \multicolumn{2}{c}{\textbf{$\Delta$t (Gyr)}} \\
    \cmidrule(lr){2-3} \cmidrule(lr){4-5}
        & \textbf{MAE} & \textbf{RMSE} & \textbf{MAE} & \textbf{RMSE} \\
    \midrule 
        Random predictor & 0.6479 & 0.7932 & 1.2306 & 1.5065 \\ 
    \midrule 
        Oriented scattering + linear & 0.4119 & 0.4953 & 0.8025 & 0.9604 \\ 
        SHWS + linear                & 0.3876 & 0.4701 & 0.7781 & 0.9360 \\
    \midrule
        SHWS + KNN regressor & 0.3539 & 0.4699  & 0.7264 & 0.9370 \\
        SHWSBic + KNN regressor & 0.3322 & 0.4498 & 0.6633 & 0.9139 \\
    \midrule
        CNN & 0.3564 & 0.4586 & 0.7206 & 0.9406 \\
        H-Net & 0.3581 & 0.4543 & 0.7345 & 0.9015 \\        
    \bottomrule 
    \end{tabular}
\end{table}

\begin{figure*}
    \begin{subfigure}{0.9\columnwidth}
        \centering
        \includegraphics[height=0.85\textwidth]{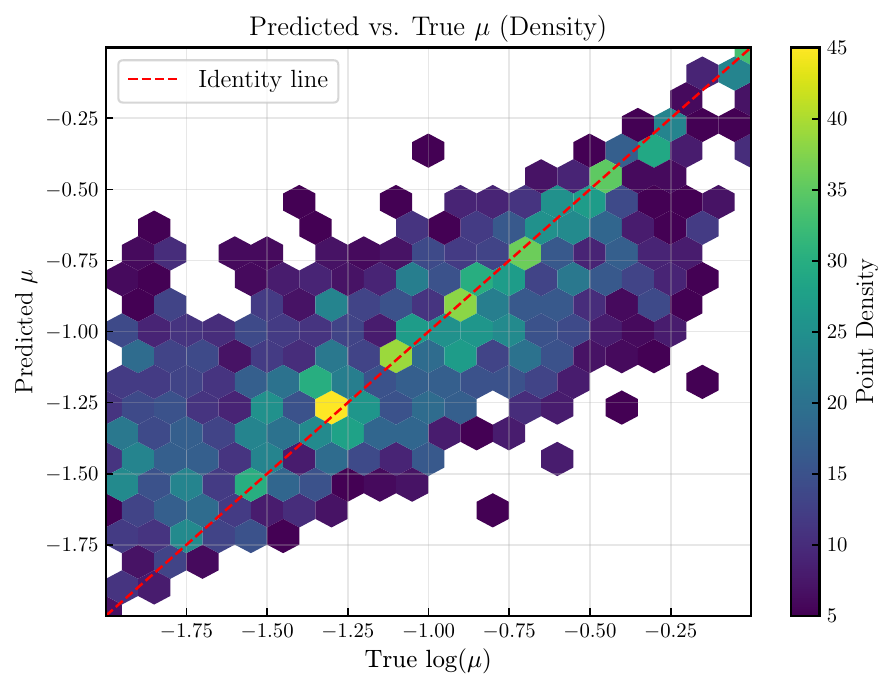}
        \caption{Log merger mass ratio}
        \label{fig:mass-density-true-pred}
    \end{subfigure}
    \hspace{0.04\textwidth} 
    \centering
        \begin{subfigure}{0.9\columnwidth}
        \centering
        \includegraphics[height=0.85\textwidth]{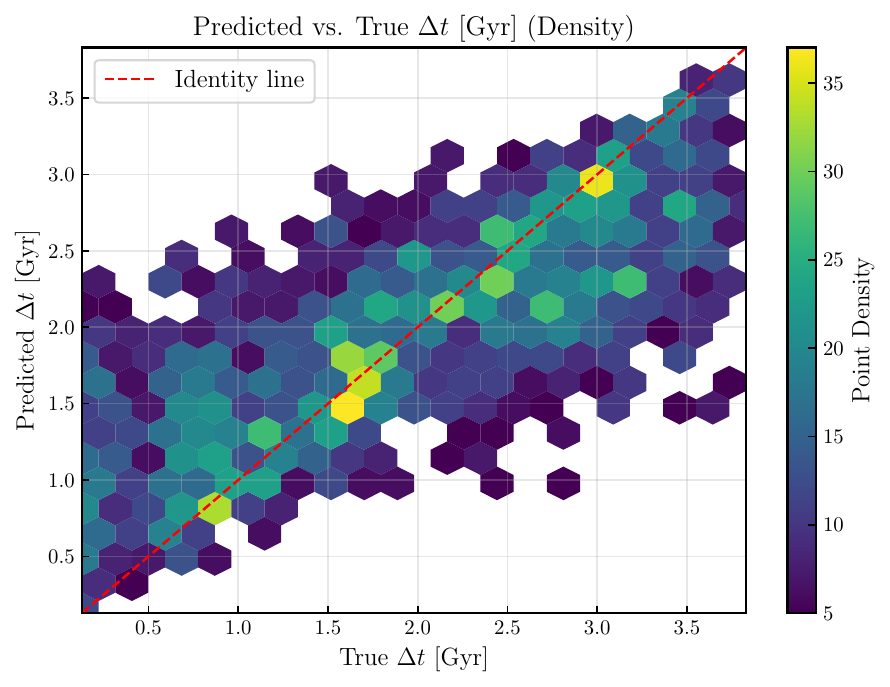}
        \caption{Time since last merger}
        \label{fig:time-density-true-pred}
    \end{subfigure}
    \caption{Results of regression on merger features using solid harmonic wavelet bispectrum. True-predicted hexbin plots using K-Nearest-Neighbours regressor on Solid Harmonic Wavelet Scattering Bicoherence features computed over noisy samples with best performance. Hexbins with more than 5 samples are shown to highlight underlying trends and remove outliers. }
    \label{fig:true-pred-regression-plots}
\end{figure*}

In Figure \ref{fig:true-pred-regression-plots} we show true-predicted plots for both merger mass ratio and time since last merger, presenting a strong correlation between true and predicted merger properties, with an uncertainty of $\sim$ 0.3 dex. The strong relationship demonstrates that our wavelet-based operator effectively encodes complex structural signatures of galaxy mergers and is able to recover detailed merger properties, albeit with uncertainties. Figure \ref{fig:true-pred-regression-plots} also shows some bias in the prediction, i.e. merger mass ratios are over/under-predicted at low/high mass ratios respectively. A similar trend is observed for time since merger. We will explore this further in Section \ref{sec:correlation_mu_deltat}, , when axemining the performance of the regression on realistic noisy data and show that this is caused by known correlation between the effects of mass ratio and time since merger on the strength of merger feautures.

\subsubsection{Impact of dataset size}

\begin{figure}
    \centering
    \includegraphics[width=\columnwidth]{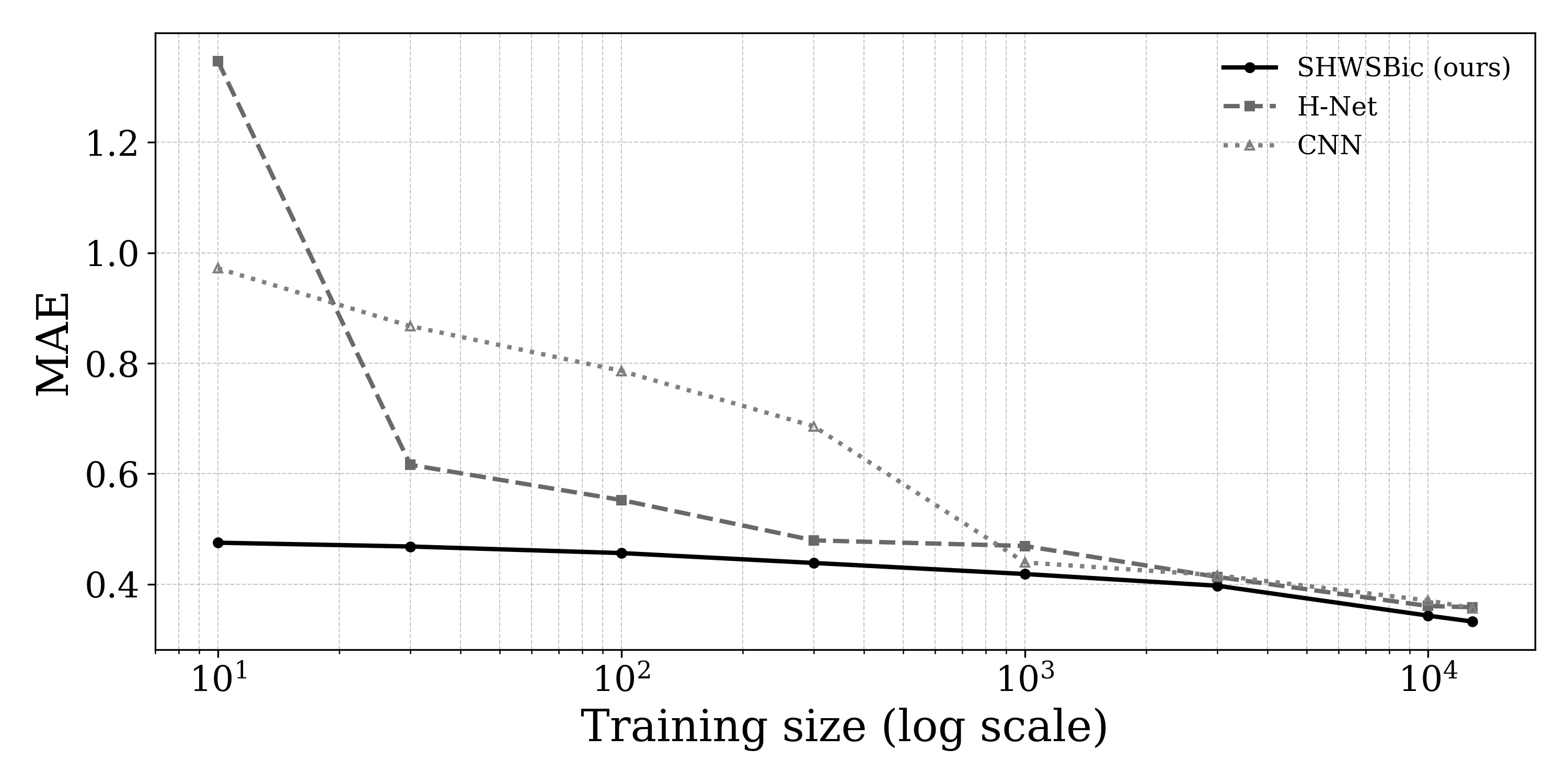}
    \caption{Required sample size for different machine learning methodologies. Comparison of mean absolute error for log mass ratio regression over increasing training set sizes. Training sets increase from 10 samples to the full set (12,868 samples) on a log scale. Our features (SHWSBic) are used to fit a KNN regressor, and were extracted from idealised samples in AB magnitude over a log scale.}
    \label{fig:dataset-size-vs-mae}
\end{figure}

Moreover, the suitability of our operator in data-scarce domains is reinforced by Figure \ref{fig:dataset-size-vs-mae}. Merger classification is often limited by the availability of suitable training samples since mergers, especially major mergers, are rare \citep[see e.g.][]{avirett-mackenzie_post-merger_2024}. In this plot, we demonstrate that the wavelet-based operator offers immediate access to discriminative features, since they are not learned from data but rather structurally encoded. Comparatively, CNNs must learn a complex hierarchy of filters in order to extract equally informative features, and so struggle with smaller training set sizes. Equivariant by design, H-Nets more efficiently learn merger dynamics in lower data regimes thanks to their abstraction of having to learn properties such as rotation-covariance -- yet  still depend on learning filters through back-propagation, and too suffer the limitations of smaller training sets. It is in data-scarce domains, therefore, that wavelet-based approaches excel. Moreover, the deep learning approaches only appear to be closing the gap towards the largest training set. 

Nonetheless, from Figure \ref{fig:dataset-size-vs-mae} it is reasonable to assume that with an increase in data such deep learning architectures could surpass the expressive capacity of our operators. Acquiring such data is both costly and time-consuming; in the case of training on simulated galaxy mergers, training sample sizes for cosmological simulations are limited by the number of snapshots, the simulation volume, as well as the rarity of mergers, expecially of equal mass ratio. Dedicated merger simulation can overcome the latter limitations, but lack the cosmological context for mergers. Simulations such as Illustris-TNG100 are commonly used \citep[e.g.][]{avirett-mackenzie_post-merger_2024, bickley_convolutional_2021, margalef-bentabol_galaxy_2024} and limit the number of mergers to the $\sim 10^3$, depending on redshift and mass ratio range used, even loosening constraints does not significantly increase the number of mergers. Similarly, human classification is also limited by the availability of classification. Our methodology can therefore outperform CNNs in common cases where training data is limited.  

\subsection{Regression on noisy data} \label{sec:regressionnoisyresults}

\begin{figure}
    \centering
    \includegraphics[width=\columnwidth]{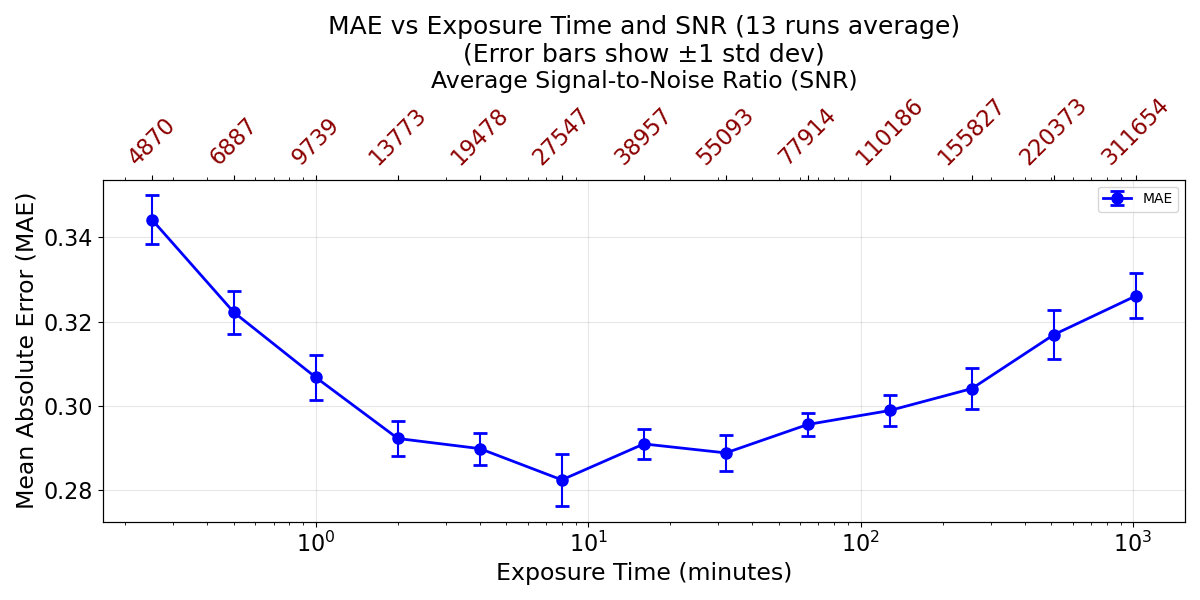}
   \caption{Noise resistance of solid harmonic wavelet bispectrum. Mean and standard deviation of the mean absolute error (MAE), averaged across ten train-test splits, as a function of exposure time $15 \times 2^{i}\,\mathrm{s}$ for $i = 0, \dots, 12$. The corresponding average signal-to-noise ratio for each exposure time is indicated at the top of the image. Images are represented in AB magnitude on a logarithmic scale.}
   \label{fig:mae-vs-exposure-time}
\end{figure}

In this section, we demonstrate the performance of our proposed approach on data mocked to resemble a low-redshift space based observation in resolution and include sky levels expected for space based nerar infrared data as well as shot noise for both sky and object. Resilience to noise is crucial for application to astronomical data. We create datasets over HST-equivalent exposure times ranging from 60 seconds to 10 hours to cover the full range from short exposures to deep imaging, in addition, we also estimate signal to noise for the integrated source, which is also cited in the top axes in Figure \ref{fig:mae-vs-exposure-time}. Not that this is the SN for the detection of the entire galaxy. Our images are therefore high-SN, high resolution and spatially sample galaxy features well. Figure \ref{fig:mae-vs-exposure-time} shows results averaged over 10 train-test splits for various noise levels when regressing merger mass ratio.

Interestingly, the operator performs equally well on noisy image samples -- up to a limit. When the exposure time decreases below $10-15$s (equivalent to an average SN $\approx 4 \times 10^3 $ for detection of entire galaxy), performance rapidly diminishes, highlighting that the signal-to-noise ratio is no longer viable for accurate feature regression. However, to an extent, the shorter the exposure time (and therefore more shot noise), the better the operator performs. Discrete and random, photon noise follows a Poisson distribution. For optical imaging application, where photon numbers are high, such noise is approximately Gaussian. Bispectral analysis is particularly powerful in identifying non-Gaussianity and therefore this noise is essentially nullified by the bispectrum over wavelet scattering coefficients. Simultaneously, sky noise and shorter exposure times result in fewer extended low surface brightness features being picked up by the telescope. Since the noise is nullified, and only the strongest features are picked up, the resultant representation is simplified in comparison to longer exposure times in which low surface brightness features become clearly detectable and thus add complexity to the representation we are aiming to learn from or encode. Moreover from Figure \ref{fig:merger-sample-increasing-exposure}, one can still identify key merger dynamics such as tidal tails with as little as 60 seconds exposure equivalent to an average SN $\approx 10^4$ for the entire object. This means that at shorter exposure times, the strongest features captured still carry informative signals with respect to merger features. Note that the current simulation is focused on low redshifts, where signal to noise is high even for short exposures, higher redshift imaging where surface brightness dimming becomes dominant. Exploring the full signal-to-noise range is beyond the scope of this current paper, however, the strong performance under addition of limited amounts of noise is encouraging for application in astrophysics.

\begin{figure*}
    \begin{subfigure}{0.9\columnwidth}
        \centering
        \includegraphics[height=0.85\textwidth]{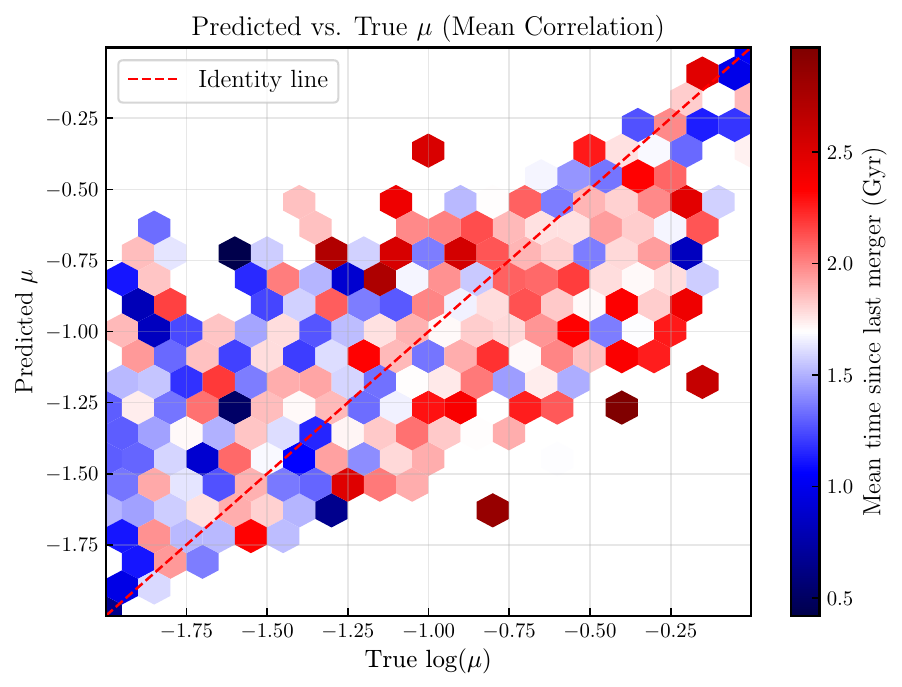}
        \caption{Log merger mass ratio}
        \label{fig:mass_correlation}
    \end{subfigure}
    \hspace{0.04\textwidth} 
    \centering
        \begin{subfigure}{0.9\columnwidth}
        \centering
        \includegraphics[height=0.85\textwidth]{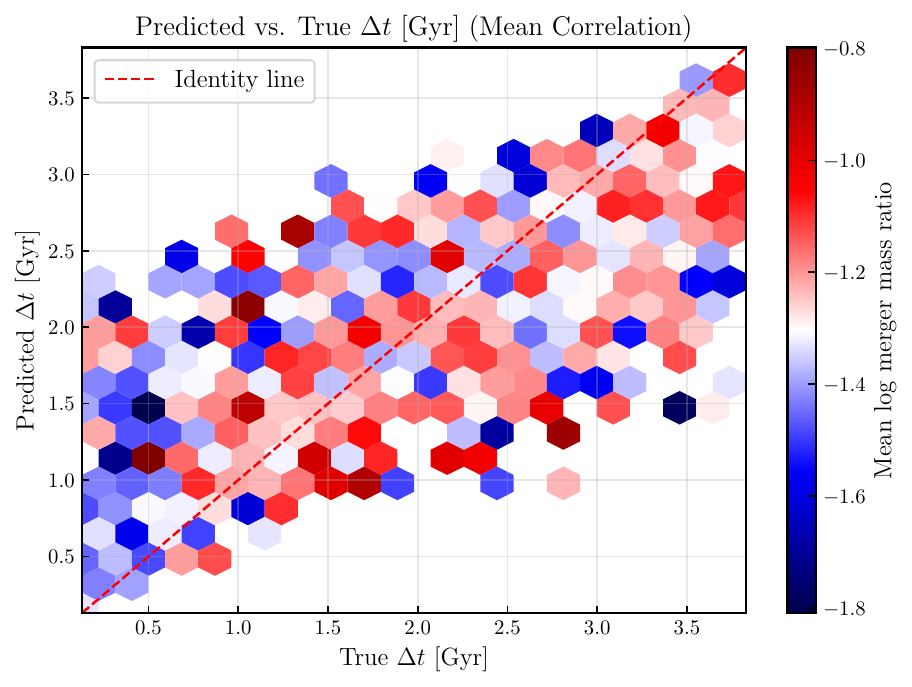}
        \caption{Time since last merger}
        \label{fig:time-correlation}
    \end{subfigure}
    \caption{Plots corresponding to Figure \ref{fig:true-pred-regression-plots} whereby the density colour mapping for each bin has been replaced with the mean of the correlative target variable. E.g. for log merger mass ratio regression, the colour represents the true mean time since last merger.}
    \label{fig:correlatory-regression-plots}
\end{figure*}

\subsection{Correlation between $\mu$ and $\Delta\textrm{t}$}
\label{sec:correlation_mu_deltat}

In this section, we examine the extent to which the correlations between the merger properties $\mu$ and $\Delta\textrm{t}$ influence the performance on merger feature regression (Figure \ref{fig:correlatory-regression-plots}); i.e. does $\Delta\textrm{t}$ affect our ability to predict $\mu$? Figure \ref{fig:correlatory-regression-plots} showed that the regression overestimated the merger mass ratio at low merger mass ratios and over-predict it at high merger mass ratio. A similar trend is observed for time since merger. A correlation in the strength of merger features between the two properties is expected, since merger features weaken both with time since merger and lower mass ratio. We therefore study the correlation between the two properties.

Figure \ref{fig:mass_correlation} shows the true and predicted mass ratio, colour-coded the time since merger. This show that our model slightly underestimates $\mu$ for longer time since last merger coalescence, and conversely trends higher for recent mergers, as expected from known trends of merger strength with both parameters. Conversely, Figure \ref{fig:time-correlation}, shows the true and predicted time since merger, colour coded by the mass ratio. When predicting $\Delta\textrm{t}$, the trend with mass ratio affecting the predicted time since merger is less distinct but still present: the model slightly overestimates $\Delta\textrm{t}$ for low mass rations, and overestimates mass ratio for low $\Delta\textrm{t}$.

This result is consistent with the expected strength of observed merger features as a function of the two parameters. For a longer time since last merger coalescence, merger signatures are expected to fade, reducing the predictive power of morphology-based features. Such features could perceptually be reminiscent of more recent, lower mass ratio mergers. Simultaneously, for more recent mergers, morphological features remain strong and could be comparable to higher mass ratio, more evolved mergers. As expected, $\mu$ and $\Delta\textrm{t}$ directly influence the ability to predict one another. This shows that for regression of merger properties, predictions for will be correlated.

\subsection{Feature Importance} \label{sec:featureimportance}

Here, we would like to explore the show how feature importance can be used to better understand classification. The features we propose are not galaxy specific but a geometric representation of rotationally covariant structure. To understand what that represents in a galaxy we demonstrate the features, on selected galaxies, ordered by importance in the regression results. In principle, this approach can be used to connect different features that drive the classification to visual features within galaxies. Although a full analysis is beyond the scope of this paper, we show here potential strength of SHWSBic in connecting classification to easily understandable features in images.
To understand which aspects of galaxy morphology the representation encodes, we perform a permutation feature importance analysis. This method performs ablations to observe which features — when removed — cause the largest change to results. We conduct this analysis on a KNN model fit to scattering bicoherence (SHWSBic) features. A more detailed and technical analysis of feature importance and how it differs from salience maps can be found in Appendix \ref{app:interpretability}.

\begin{figure}
    \centering
    \includegraphics[width=\columnwidth]{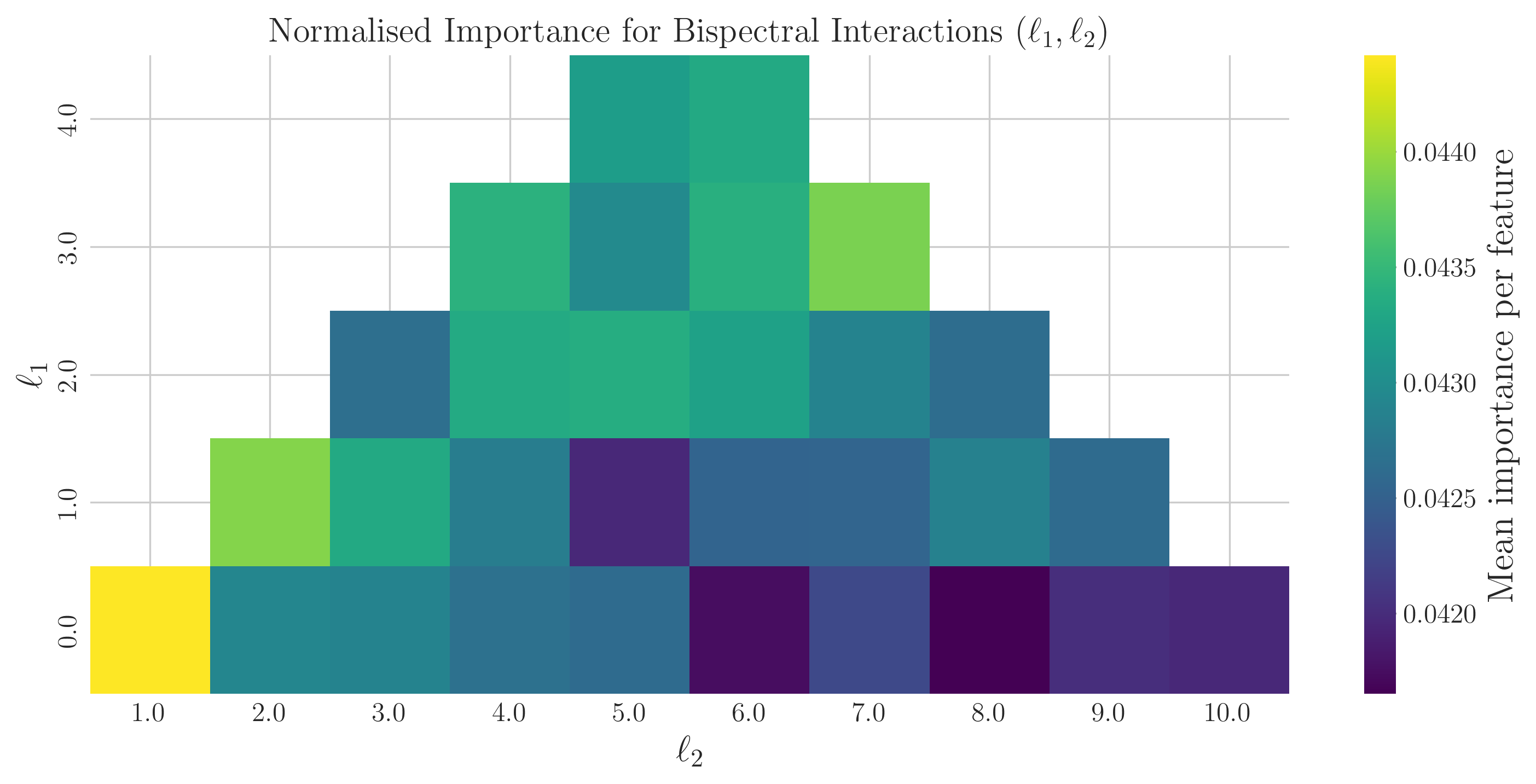}
    \caption{Mean feature importance of bispectral interactions. Note that due to the symmetry of the bispectrum, we only compute for $\ell_1, \ell_2 \in [0, L]$ where $\ell_1 < \ell_2$.}
    \label{fig:feature-importance-bispectral-interactions}
\end{figure}

\begin{figure}
    \centering
    \includegraphics[width=0.8\columnwidth]{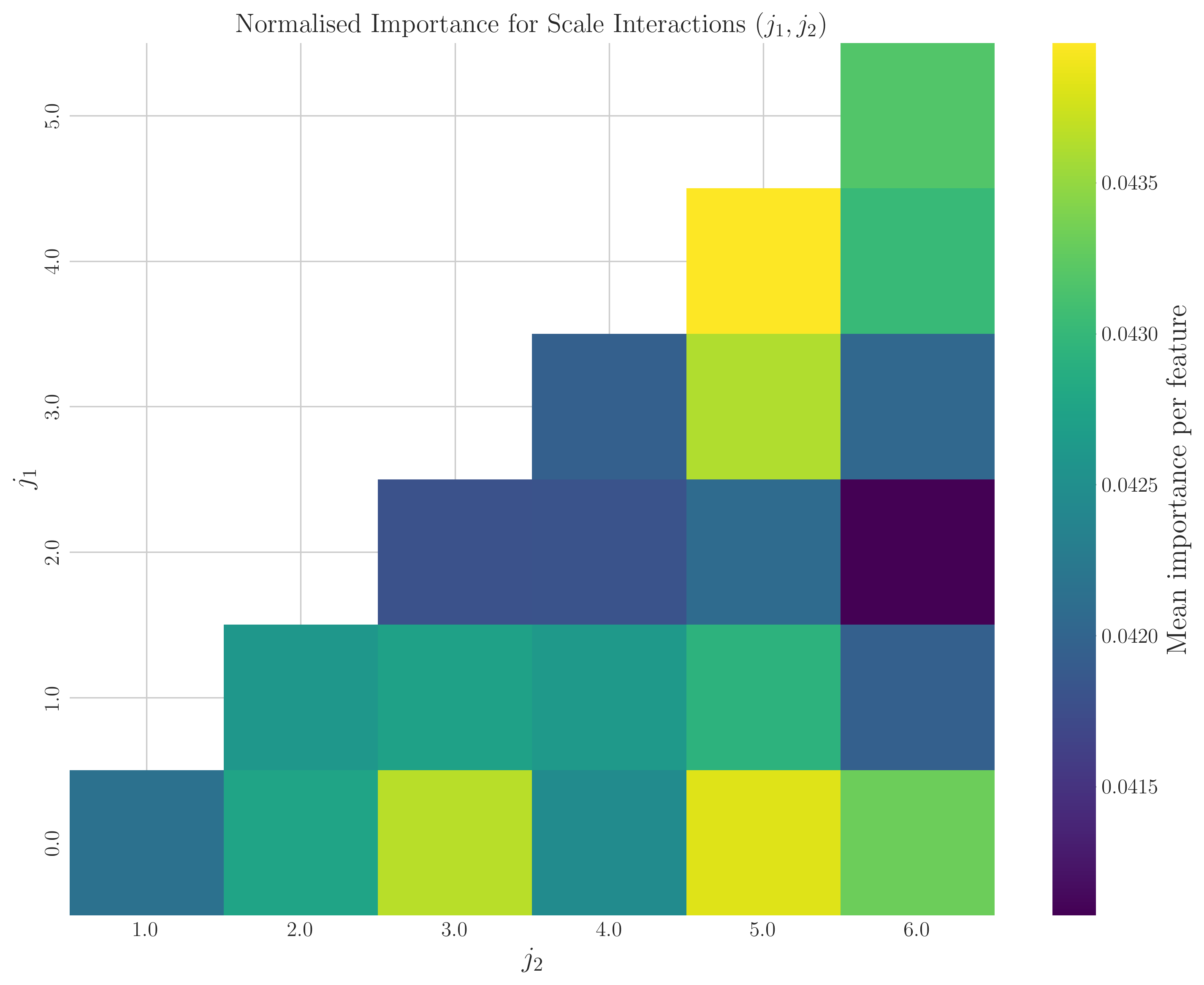}
    \caption{Mean feature importance of scale interactions for second-order coefficients.}
    \label{fig:feature-importance-scale-interactions}
\end{figure}

Across scales, angular frequencies, and their interactions, feature importance is distributed uniformly — suggesting an economical representation with no redundant features. Figure \ref{fig:feature-importance-bispectral-interactions} shows the normalised contributions of different angular frequency combinations $(\ell_1, \ell_2)$ used in the bispectrum calculation. Each cell represents the importance of a particular interaction between two angular frequencies, where $\ell_1 < \ell_2$ due to the symmetry of the bispectrum, and $\ell_1 + \ell_2 \leq 10$ since we only compute coefficients up to $\ell = 10$. The most and least significant bispectral features differ by approximately $5\%$, suggesting that all combinations are similarly informative for prediction. Similarly, Figure \ref{fig:feature-importance-scale-interactions} shows scale combinations $(j_1, j_2)$ for the second-order scattering coefficients, where $j_1$ corresponds to the scale of the first-order response and $j_2$ to the scale of the second wavelet operator applied to it. Their significance is also uniformly distributed. This indicates that regression on merger features integrates information from multi-scale morphological structures broadly, rather than relying on specific feature subsets.

\begin{figure}
    \centering
    \includegraphics[width=\columnwidth]{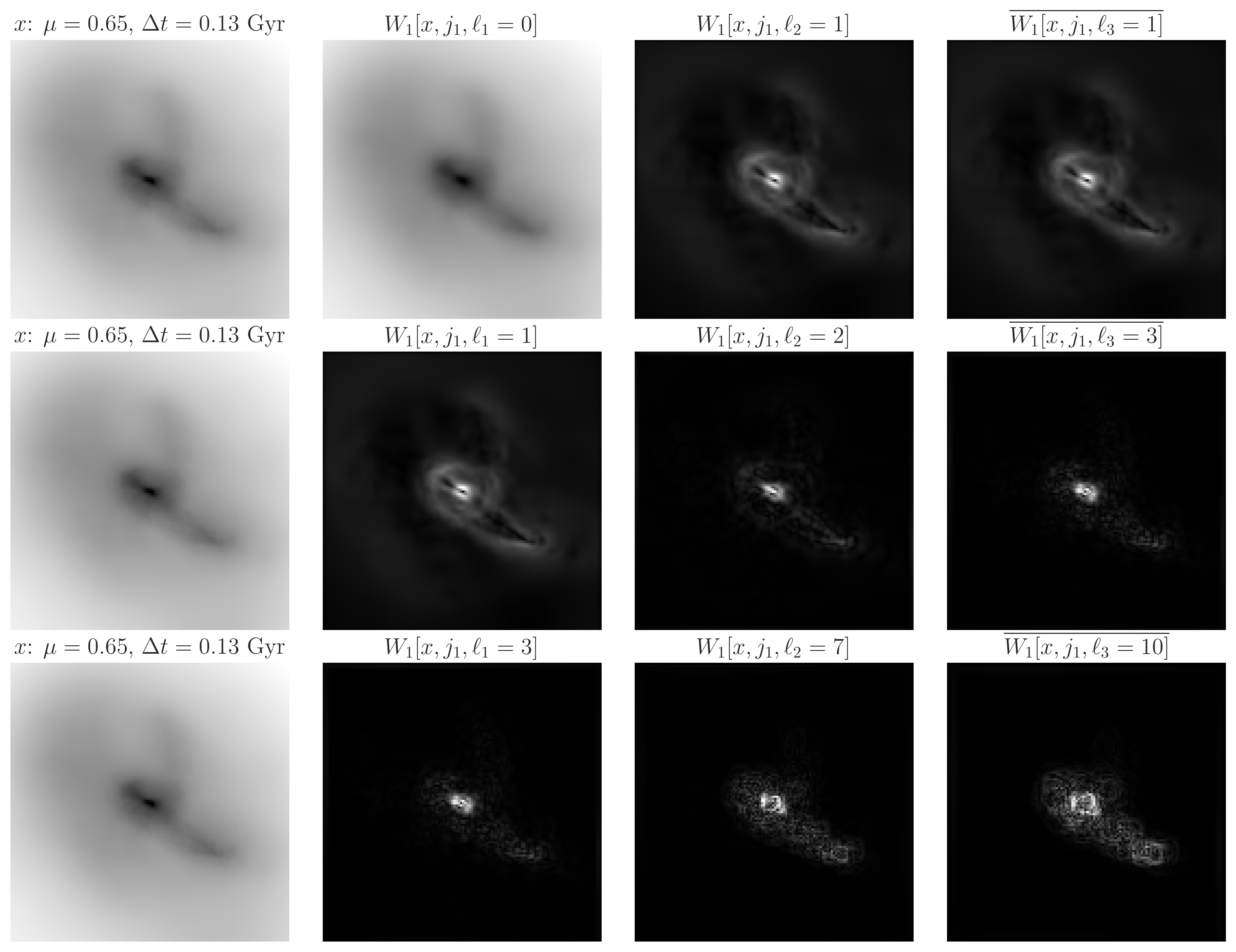}
    \caption{Real components of features extracted by the solid harmonic wavelet transform for the most important $(\ell_1, \ell_2)$ interactions, as depicted by Figure \ref{fig:feature-importance-bispectral-interactions}. Visualisations for more galaxies can be found in Appendix \ref{apdx:more-bispectral-feature-interactions}.}
    \label{fig:features-extracted-bispectral-interactions}
\end{figure}

To illustrate what the representation captures at different bispectral interactions, we visualise features extracted for representative $(\ell_1, \ell_2)$ pairs in Figure \ref{fig:features-extracted-bispectral-interactions}. At $(\ell_1=0, \ell_2=1)$, the transform captures the co-occurrence of smoothed structure (since $\ell_1=0$ simplifies to a Gaussian) with directional features, emphasising spiral arms. At $(\ell_1=1, \ell_2=2)$, the bispectrum encodes interactions between two similar levels of directional features. At higher angular frequencies such as $(\ell_1=3, \ell_2=7)$, the transform captures relationships between low-level rotational symmetries and complex directional patterns, encoding structures such as tidal tails, asymmetries on different physical scales and double-nuclei. These interactions quantify the co-occurrence of different rotational symmetries in galaxy morphology.

\begin{figure}
    \centering
    \includegraphics[width=\columnwidth]{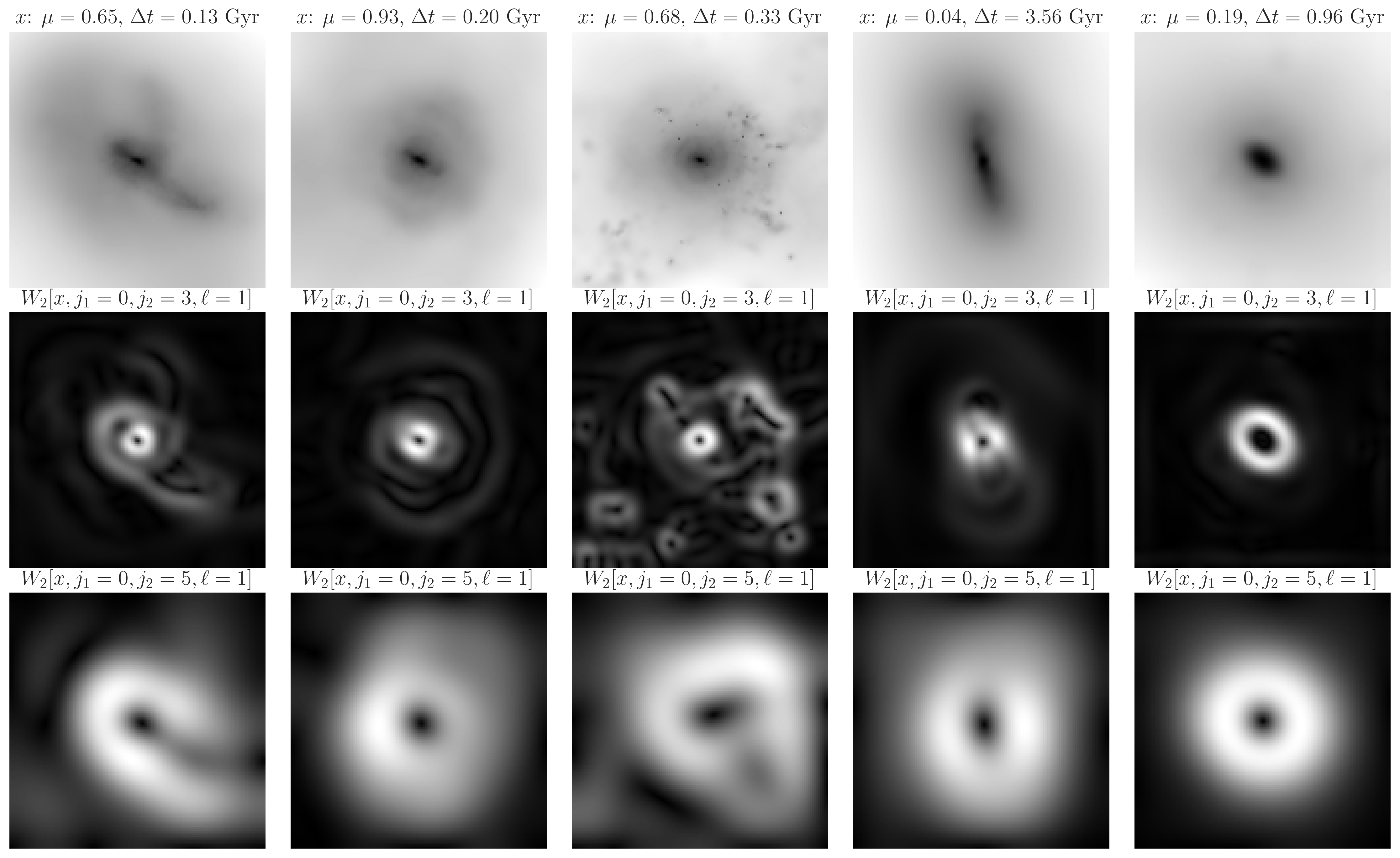}
    \caption{Real components of features extracted by second-order solid harmonic wavelet scattering for important $(j_1, j_2)$ pairs, as highlighted by Figure \ref{fig:feature-importance-scale-interactions}.}
    \label{fig:features-extracted-second-order}
\end{figure}

We apply the same approach to second-order scattering features produced from scale interactions. As with angular frequencies, different scale pairings have similar contributions to predictive performance. To illustrate the morphological structures captured, we visualise representative $(j_1, j_2)$ pairs in Figure \ref{fig:features-extracted-second-order}. Distant scale pairings such as $(j_1=0, j_2=3)$, $(j_1=0, j_2=5)$ and $(j_1=0, j_2=6)$ capture hierarchical relationships between fine and coarse features, while adjacent larger scales $(j_1=4, j_2=5)$ encode extended structure such as spiral arms. Such features allow for discrimination between global structures in merger samples, complementing the finer first-order features depicted in Figure \ref{fig:features-extracted-bispectral-interactions}. This reinforces the need for not only multi-scale analysis, but the modelling of increasingly complex interactions — such as those between scales.

Figure \ref{fig:features-extracted-second-order} affirms that key second-order interactions encode broader structures that are present across scales (e.g., spiral arms in column 1). Such features allow for discrimination between global structures in  merger samples, offering an enhanced representation on top of the finer first-order features as depicted in Figure \ref{fig:feature-importance-bispectral-interactions}. It reinforces the need for not only multi-scale analysis, but the modelling of increasingly complex interactions -- such as those between scales.

These results demonstrate that the solid harmonic wavelet scattering bispectrum encodes physically meaningful morphological structures — spiral arms, central bulges, and hierarchical cross-scale features — which the KNN regressor leverages in equal measure to predict merger properties.

\section{Discussion}

Here, we have shown how a solid harmonic wavelet bispectrum approach techniques can be used to improve the performance of current state of the art methodology (see Section \ref{sec:mergerclassificationresults}), achieving similar or better accuracy while being resistant to noise (see Fig. \ref{fig:mae-vs-exposure-time}) and requiring smaller datasets (see Fig. \ref{fig:dataset-size-vs-mae}). We note here that while the full sample size we use is comparable to similar studies, Fig. \ref{fig:dataset-size-vs-mae} shows that the method also performs very well with much smaller sample sizes, in contract to other methods. We have also shown that our novel methodology, can be used to regress both time since mergers and mass ratio, expanding on previous work in the literature regressing on time since merger \citep{pearson2024} (see Figure \ref{fig:true-pred-regression-plots}). This methodology therefore represents an alternative to CNNs, which better performance for small datasets (Fig. \ref{fig:dataset-size-vs-mae}) and  able to perform regression on merger properties, with performance exceeding traditionally used methods.

Machine learning has been used in the literature for morphological classification of galaxies \citep[e.g.][]{huertas-company_galaxy_2024}, both for general galaxy morphologies \citep[e.g.][]{dominguez-sanchez2018,dominguez-sanchez2019,cheng_optimizing_2020} and mergers \citep[e.g.][]{bickley_convolutional_2021,wavelet_scattering_radio_galaxy_morphologies_tolley,margalef-bentabol_galaxy_2024}. The commonly used method has been CNNs, in some case bayesian CNNs \citep{walmsley_galaxy_2020} and inclusion of active learning techniques \citep{walmsley_zoobot2023}. The importance of realism, as well as the impact of lower quality data has been analysed in several studies \citep{pearson2019, bottrell_deep_2019}. Transfer learning to adapt to different datasets \citep{dominguez-sanchez2019}. Contrastive learning has also been used to account for contamination \citep{huertas-company_contrastive2023}. Accuracies for general morphological features are high and can be well in excess of 90\% \citep[e.g.][]{dominguez-sanchez2018}, but can be significantly lower for the detection of mergers, which show weaker and less well defined features \citep{margalef-bentabol_galaxy_2024,omori2023}. Classification of mergers using CNNs typical achieves accuracies $\sim 80-85 \%$, with exact accuracies depending on image quality and merger samples \citep{margalef-bentabol_galaxy_2024, avirett-mackenzie_post-merger_2024, omori2023}. While this accuracy is relatively high, the fact that mergers are rare can lead to overall low purity in merger samples \citep{lambrides_mergerstat_2021}. CNN-based methods have been used to identify mergers in large image samples as well as to study the impact of mergers on other physical galaxy properties, such as AGN activity. 

One of the limitations of common methodologies in the literature is that most studies have used binary classifiers, i.e. mergers and non-mergers. However, in simulations, the impact of mergers on galaxies depends on properties of the mergers, including the mass ratio and time since merger \citep{cox2008,Scudder2015}. Several studies have therefore attempted to derive more nuanced merger classification. \citet{margalef-bentabol_galaxy_2024} presented a merger challenge and showed that while all participants performed well on binary classification, performance dropped on multiple labels (to $\sim$ 50-60 \%). Since then, other studies have attempted classification using several stages \citep{ferreira2024, de_graaff_classifying_2025}.

\citet{ferreira2024} employed a multi-stage classification of mergers, which improves the performances compared to CNNs significantly to $\sim$95\%. However, compared to our methodology, this approaches has more free parameters, and therefore requires larger training sets. Similarly, \citet{de_graaff_classifying_2025} use Zoobot \citep{walmsley_zoobot2023} to train a merger sample similar to the one used here into multiple labels and also uses a multi-stage classification, similar to \citep{ferreira2024}, but found a one-stage setup to outperforms the two stage setup. Specifically, \citet{de_graaff_classifying_2025} found the highest accuracy for pre-mergers, which show double-nucleated structures, as opposed to more complex post-mergers.

\citet{ferreira2026} used MUMMI to classify merger timescales by creating bins in time scale since merger traines on simulations and show that their model can predict timescale. The model used in this paper performs well at classification, but uses only timescale, rather than timescale and mass ratio. The model used in this paper also has a large number of free parameters, similar to model tested here, so our model will require smaller sample sizes.

\citet{pearson2024} also studied timescale since merger, using a CNN, they found that uncertainties in the time since merger where comparable to the time between snapshots, in their work, the CNN outperformed other methods. This uncertainty is lower than presented here, but includes only major mergers, as well as includes the more easily identifiable pre-merger stages, where the presence of binary sources makes mergers more easily identifiable. Given that our method outperforms their leading model of CNNs. \citep{pearson2025} applies such methodology to study the impact of mergers on galaxy properties, showing the potential application of regression models to understand the physics of mergers.

Therefore, while CNNs have shown strengths in analysing merger features, they perform less well on more complex tasks, such as merger features and have so far only shown application in classification of mergers based on a small number of labels.

The regression methodology applied here can be applied to large galaxy samples from e.g. Euclid or LSST to derive distributions of merger properties for large samples. While uncertainties on individual galaxies are high, regressed merger features can be used to determine the impacts of mergers of different mass ratios on e.g. star formation and black hole growth, as well as compare the timing of black hole accretion during mergers to simulations.

\section{Conclusions}

In this work, we performed regression on continuous merger features (both time since merger and mass ratio) from morphological image data, with applications in large-scale astronomical surveys using the solid harmonic wavelet bispectrum. We show for the first time regression on both mass ratio and time since mergers, expanding on previous enhanced classification of mergers \citep{pearson2024,ferreira2026} We begin by demonstrating the applicability of wavelet-based techniques in comparison to prior approaches on merger classification (Section \ref{sec:mergerclassificationresults}), before progressing to regression on merger properties (Section \ref{sec:regressionidealisedresults}, \ref{sec:regressionnoisyresults}). We demonstrate the strengths of these methodologies in data-scarce regimes. Our results can be summarized as follows:

\begin{itemize}
    \item To analyse the performance of different approaches on galaxy classification, we select a sample of galaxy mergers from IllustrisTNG, with a wide range of galaxy mass ratios and time since mergers. We analyse both idealised and high-SN high-resolution images. (Section \ref{sec:mergerdata})
    \item We compare several wavelet-based methodologies — Morlet wavelets with linear regression, solid harmonic wavelets with linear regression, and the solid harmonic wavelet bispectrum with both linear regression and KNN regression — against conventional CNNs and Geometric Deep Learning models (Section \ref{sec:mergerclassificationresults}).
    \item Although CNNs slightly outperform solid harmonic wavelets in simple binary classification tasks (Section \ref{sec:mergerclassificationresults}), solid harmonic wavelet bispectrum methodologies allow regression on merger features with uncertainties $\sim 0.33$ on $\textrm{log}(\mu$) and $\sim 0.66$ Gyr on $\Delta\textrm{t}$ on idealized images (Fig. \ref{fig:correlatory-regression-plots}). We further show that some of the scatter in the predictions is driven by the known dependence of merger features on the mass ratio and the time since the merger (Fig. \ref{fig:features-extracted-bispectral-interactions}). This method therefore opens up the possibility to derive physical merger properties, rather than classifying galaxies into broad categories.
    \item We demonstrate the robustness of the proposed method to noise using a wide range of signal-to-noise conditions (Section \ref{sec:regressionnoisyresults}). Suggesting the approach would be a promising candidate methodology for more realistic regimes.
    \item In terms of data efficiency, we find that solid harmonic wavelet bispectrum features with k-nearest-neighbour regression perform extremely well for small sample sizes, due to the minimal number of trainable parameters required (Figure \ref{fig:dataset-size-vs-mae}).
\end{itemize}

We therefore show that the solid harmonic wavelet bispectrum offers an appealing alternative to conventional deep learning techniques, with better performance especially at small sample sizes and a transparent functional methodology (Section \ref{sec:featureimportance}). In the future, such models can be applied to new large datasets (e.g. LSST, Euclid) to better understand the impact of mergers on galaxy evolution.

\section*{Acknowledgements}

We thank both reviewers for detailed comments which helped improve the manuscript. Funding acknowledgements: This project has received funding from the European Union’s Horizon 2020 research and innovation programme under the Marie Skłodowska-Curie grant agreement no. 860744.
This work made use of Astropy:\footnote{http://www.astropy.org} a community-developed core Python package and an ecosystem of tools and resources for astronomy \citep{astropy:2013, astropy:2018, astropy:2022}. 
This work made use of Kymatio:\footnote{https://www.kymat.io/} an open-source implementation of the wavelet scattering transform in Python \citep{kymatio}.
This research made use of Hex, the GPU Cloud in the Department of Computer Science at the University of Bath

%%%%%%%%%%%%%%%%%%%%%%%%%%%%%%%%%%%%%%%%%%%%%%%%%%
\section*{Data Availability}

The galaxy image data are taken from IllustrisTNG, the data are publicly available through this \href{https://www.tng-project.org/data/}{webpage}. Upon publication, code to create and analyse the data will be made available through GitHub.



\bibliographystyle{mnras}
\bibliography{example} % if your bibtex file is called example.bib

%\begin{thebibliography}{99}
%\end{thebibliography}

%%%%%%%%%%%%%%%%%%%%%%%%%%%%%%%%%%%%%%%%%%%%%%%%%%

%%%%%%%%%%%%%%%%% APPENDICES %%%%%%%%%%%%%%%%%%%%%

\appendix

\section{Background}
\label{apdx:background-knowledge}
In this section, we offer introductions to multiple concepts which are central to the core theory in this paper: wavelet scattering and higher-order statistics.

\subsection{Solid Harmonic Wavelets}
\label{apdx:solid harmonic wavelets}

\begin{figure}
    \centering
    \includegraphics[width=\columnwidth]{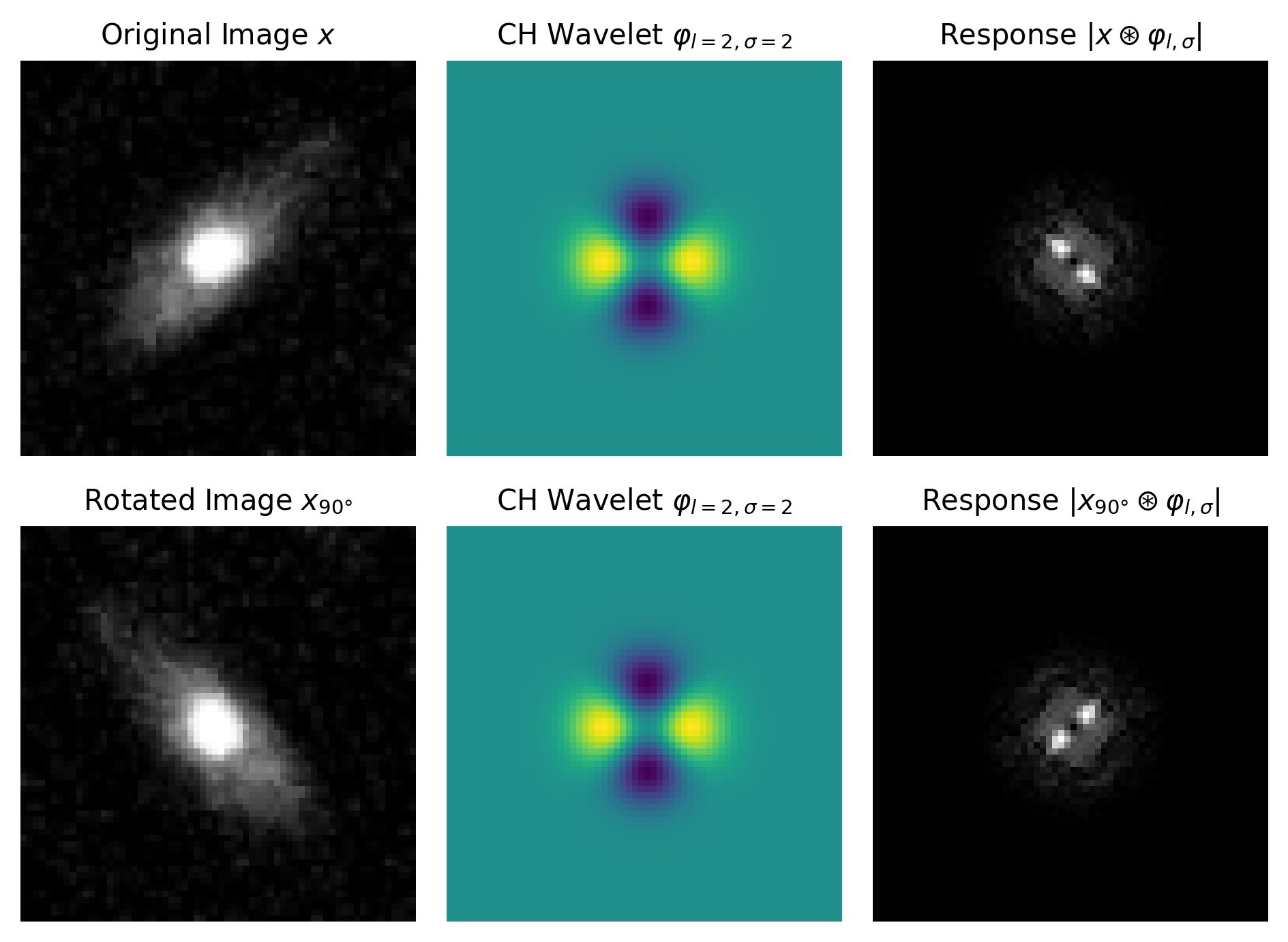}
   \caption{Demonstration of rotation-covariance property of solid harmonic wavelet transform. Rotated signals transform predictably.}
   \label{fig:rotation-covariance}
\end{figure}

Solid Harmonic Wavelets \citep{eickenberg2017solid} provide an orthogoncal basis with a rotation covariance property that is central to developing rotation invariant features.
Rotating the input signal results in a corresponding rotation on the extracted features, as can be visualised in Figure \ref{fig:rotation-covariance}.
The integrals of the outputs of Solid Harmonic Wavelets, third column of Figure \ref{fig:rotation-covariance}, are going to be invariant to rotation.

\subsection{Oriented Wavelets}
\label{apdx:oriented_wavelets}

\begin{figure}
    \includegraphics[width=\columnwidth]{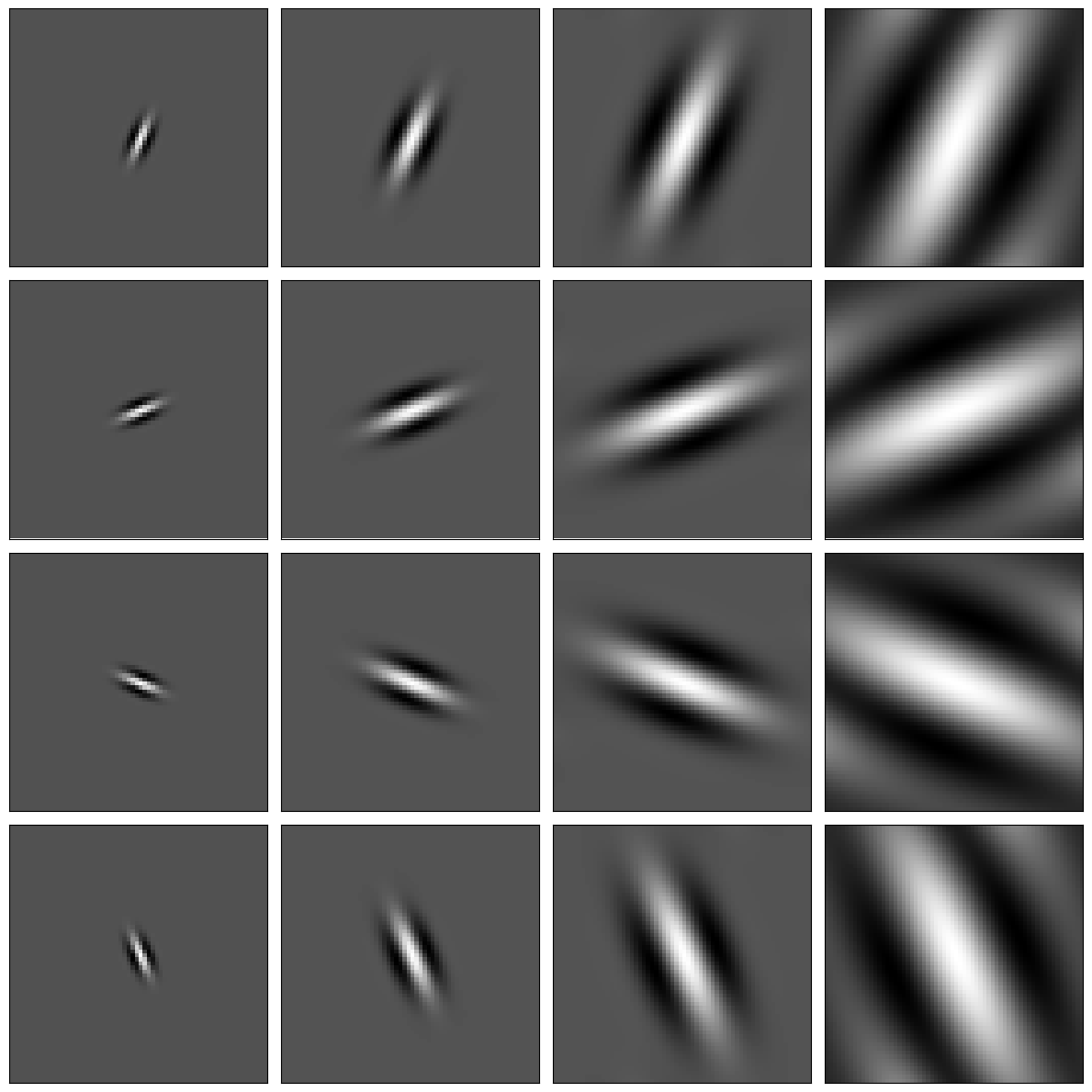}
    \centering
    \caption{Oriented Morlet wavelets. Top-to-bottom: sampled orientation $\theta$  increases. Left-to-right: scale $j$ increases.}
    \label{fig:morlet-wavelet}
\end{figure}

\begin{figure}
    \centering
    \includegraphics[width=\columnwidth]{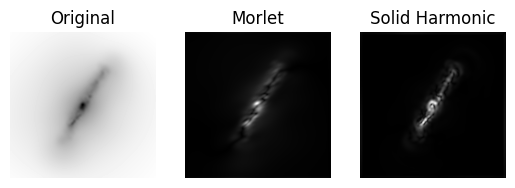}
   \caption{Comparison of features extracted by Morlet and Solid Harmonic wavelet transforms on an idealised sample on a log scale over AB magnitude.}
   \label{fig:morlet-vs-solid-harmonic-wavelet-transform}
\end{figure}

Most scattering implementations \citep{brunamallat2012scattering,Oyallon_2019_hybrid,kymatio} consider oriented wavelets, such as the Morlet wavelets visualized in Figure \ref{fig:morlet-wavelet}. The responses of Morlet wavelets when compared to Solid Harmonics appear to highlight image structure only when that structure is aligned with the wavelet direction, see for instance Figure \ref{fig:morlet-vs-solid-harmonic-wavelet-transform}, while solid harmonic wavelets appear to preserve structure regardless of orientation.

\subsubsection{Morlet Wavelets} \label{sec:morlets}

Morlet wavelets can be thought of as localised sine waves in 2D -- a smooth filter capable of detecting edges and textures. They are defined as the product of a Gaussian window with scale $\sigma$ by a sine wave with frequency $\xi$ \citep{brunamallat2012scattering}:
\begin{equation}
\psi_{\sigma, \xi}(u) = C_1 \left(e^{i u \cdot \xi} - C_2\right) e^{- \frac{|u|^2}{2\sigma^2}}
\end{equation}
where $u = (x, y)$, and $C_1$, $C_2$ are normalisation constants. The Gaussian ensures the wavelet is spatially localised, and can be configured over various dilations for multi-scale analysis. Features can be extracted using the two-dimensional discrete wavelet transform of a signal $x$ using Morlet wavelets at scale $\sigma$ and frequency $\xi$, which is given by the convolution of the signal with the wavelet:
\begin{equation}
W[x, \sigma, \xi] = x \star \psi_{\sigma, \xi} \label{morlet convolution}
\end{equation}

\subsubsection{Morlet Wavelet Scattering} \label{sec:morletscattering}

Wavelet scattering networks \citep{brunamallat2012scattering} cascade discrete wavelet transforms with modulus nonlinearities and spatial averaging. This is similar to a CNN but with predefined filters, offering stability to deformation in addition to structurally imposing translation-invariance. It is computed over successive orders, whereby each layer progressively encodes higher-order interactions:
\begin{align}
    & S_0[x, J] = x \star \phi_J \label{eq:zeroth-order} \\
    & S_1[x, \xi_1, j, J] = \left| x \star \psi_{j, \xi_1} \right| \star \phi_J \label{eq:first-order} \\
    & S_2[x, \xi_1, \xi_2, j, J] = \left| \, \left| x \star \psi_{j, \xi_1} \right| \star \psi_{j, \xi_2} \, \right| \star \phi_J \label{eq:second-order}
\end{align}
where $\phi_J$ is a low-pass filter of the largest scale $J$; e.g. a Gaussian. Increasing orders capture interactions between features across scales, and recover high-frequency variability lost through the modulus and averaging of earlier layers. Central to merger analysis, the ability to capture correlations across scales means features like spirals and rings will emerge strongly in second-order coefficients.
%w
Oriented wavelets \citep{sifre_oriented_wavelets_texture_discrimination} are a natural extension, obtained by sampling Morlet wavelets over a discrete set orientations:
\begin{align} 
\psi_\theta(x) = \psi(r_{-\theta}x),
\end{align} 
where $r_{-\theta}$ denotes a rotation of $\theta$. This approach encodes directional features which enables the design of rotation-invariant features through spatial averaging. We make use of the open-source Kymatio framework \citep{kymatio} for oriented Morlet wavelet scattering. Oriented Morlet wavelets are visualised in Figure \ref{fig:morlet-wavelet}.

\subsection{Higher-order spectra}

Second-order statistics and their Fourier domain equivalent, the power spectrum, are widely used in signal processing to encode the distribution of energy across frequencies. However, they are incomplete and struggle to characterise minor variability with respect to morphological structure. Further, it discards phase information by construction - the power spectrum at frequency $f$ is the product of the Fourier transform and its complex conjugate (Eq. \eqref{eq:power-spectrum}).
\begin{equation}
    P(f) = F(f) \overline{F(f)} \label{eq:power-spectrum}
\end{equation}
Higher-order spectra build on these limitations, extending into third-order statistics - the bispectrum. This captures quadratic phase coupling between frequency components $f_1, f_2$ and $f_{1+2}$ (Eq. \eqref{eq:bispectrum}), encoding nonlinearities not present in the power spectrum while preserving relative phase relationships.
\begin{equation}
    B(f_1, f_2) = F(f_1) F(f_2)  \overline{F(f_{1+2})} \label{eq:bispectrum}
\end{equation}
However, the magnitude of the bispectrum depends on the underlying power at each frequency and is not solely indicative of coupling strength. To separate the contribution of phase alignment from overall signal energy, one defines the bicoherence as a normalised bispectrum (Eq. \eqref{eq:bicoherence}).
\begin{equation}
    \text{Bic}(f_1, f_2) 
    = \frac{\left| F(f_1) F(f_2) \overline{F(f_1 + f_2)} \right|}{\sqrt{ |F(f_1)|^2 |F(f_2)|^2 |\overline{F(f_1 + f_2)}|^2 }} \label{eq:bicoherence}
\end{equation}
This normalisation bounds results to the interval $[0, 1]$ with values closer to 1 indicating stronger phase coupling between the two frequencies.

\subsection{Solid Harmonic Wavelet Bispectrum Covariance}
\label{apdx:shwb-covariance-proof}
To establish the covariance of the solid harmonic wavelet bispectrum (SHWB) under rotations in Eq. 
\ref{bispectrum covariance proof}, 
we begin by considering the rotational behavior of the solid harmonic wavelet transform. For a planar rotation $R_\gamma$, the wavelet coefficients transform as shown in Eq. \ref{eq:signal-rotation-commutes}:
\begin{equation*}
W_{j,\ell}(R_\gamma x) = e^{i\ell \gamma} R_\gamma W_{j,\ell}(x),
\end{equation*}

where $R_\gamma$ denotes the spatial rotation operator acting on the coordinates, and the phase term $e^{i\ell \gamma}$ arises from the equivariance of solid harmonics under rotation of the input signal.

We substitute Equation \ref{eq:signal-rotation-commutes} into the definition of the SHWB:
\begin{equation*}
\text{SHWB}[x, j, \ell_1, \ell_2] = W_{j, \ell_1}(x) \cdot W_{j, \ell_2}(x) \cdot \overline{W_{j, \ell_1 + \ell_2}(x)},
\end{equation*}
Applying a rotation $R_\gamma$ to the input $x$ and using Equation \ref{eq:signal-rotation-commutes}, we have:
\begin{align}
&\text{SHWB}[R_\gamma x, j, \ell_1, \ell_2]
\nonumber \\
& = W_{j, \ell_1}(R_\gamma x) \cdot W_{j, \ell_2}(R_\gamma x) \cdot \overline{W_{j, \ell_1 + \ell_2}(R_\gamma x)} \nonumber \\
&= e^{i\ell_1\gamma} R_\gamma W_{j, \ell_1}(x) \cdot e^{i\ell_2\gamma} R_\gamma W_{j, \ell_2}(x) \cdot \overline{e^{i(\ell_1 + \ell_2)\gamma} R_\gamma W_{j, \ell_1 + \ell_2}(x)} \nonumber \\
&= e^{i(\ell_1 + \ell_2 - (\ell_1 + \ell_2))\gamma} R_\gamma \left(W_{j, \ell_1}(x) \cdot W_{j, \ell_2}(x) \cdot \overline{W_{j, \ell_1 + \ell_2}(x)}\right) \nonumber \\
&= R_\gamma \text{SHWB}[x, j, \ell_1, \ell_2]. \label{bispectrum covariance proof full}
\end{align}
Thus, the SHWB is \emph{covariant} to rotations. Crucially, this covariance hinges on the angular momentum conservation condition $\ell_3 = \ell_1 + \ell_2$, which ensures that the global phase factor $e^{i(\ell_1 + \ell_2 - \ell_3)\gamma}$ cancels. This cancellation is what distinguishes the SHWB from classical wavelet scattering: the phase information is preserved without requiring a modulus to enforce rotational invariance.

Since the underlying wavelet convolutions are also covariant to translations \citep{mallat2012groupinvariantscattering}, the SHWB inherits translation covariance. Full \emph{roto-translation invariance} is then achieved by applying a suitable \emph{reduction operator}, such as spatial averaging over the domain.

This formalism justifies the invariance properties of SHWB and illustrates how higher-order interactions can retain geometric structure without sacrificing phase information, offering a more expressive alternative to classical scattering transforms.

Finally, we note that the solid harmonic wavelets $\psi_{\ell,j}$, producing the responses $W_{j,\ell}(x)$, are orthogonal across different scales $j$ and angular frequencies $\ell$. As a result, the SHWB generates a set of largely independent invariants. This orthogonality ensures that while irrelevant variations (e.g., due to transformations) are removed, the essential discriminative information in the signal is largely preserved. Consequently, the SHWB forms a rich invariant representation that captures higher-order geometric structure without discarding informative phase content.

\section{Deep Learning Model Architectures}
\label{apdx:deep-learning-architectures}
This appendix provides detailed architectures for the deep learning models used in this work.

\subsection{Merger classification}
Table \ref{tab:hybrid_cnn_architecture} shows the architecture of the hybrid scattering-CNN model used for binary merger classification (Section \ref{sec:mergerclassificationresults}). The model first applies a fixed Morlet wavelet scattering transform with $L=8$ orientations and maximum scale $J=2$, producing 81 scattering coefficients per colour channel. These are reshaped and passed through a series of convolutional layers with batch normalisation and dropout for regularisation. The architecture uses grouped and depthwise convolutions to reduce parameter count while maintaining representational capacity. The final dense layers produce a two-class output for merger/non-merger classification. The total of 171,560 trainable parameters represents an almost seven-fold reduction compared to the standalone CNN from \cite{avirett-mackenzie_post-merger_2024}.

\begin{table}
\centering
\caption{Architecture of the hybrid scattering-CNN model. The wavelet scattering transform is fixed and non-trainable; all subsequent layers are trainable.}
\small
\begin{tabular}{|l|c|c|}
\toprule
\textbf{Layer} & \textbf{Output Shape} & \textbf{\# Parameters} \\
\midrule
Input Image & $(3, 36, 36)$ & -- \\
\midrule
Morlet Wavelet Scattering ($L=8, J=2$) & $(3, 81, 9, 9)$ & 0 \\
Reshape & $(243, 9, 9)$ & 0 \\
\midrule
BatchNorm-1 & $(243, 9, 9)$ & 486 \\
\midrule
Conv2D-1 ($1 \times 1$) & $(64, 9, 9)$ & 15,616 \\
BatchNorm-2 & $(64, 9, 9)$ & 128 \\
ReLU-1 & $(64, 9, 9)$ & 0 \\
Dropout-1 & $(64, 9, 9)$ & 0 \\
\midrule
Grouped Conv2D-2 ($3 \times 3$) & $(128, 7, 7)$ & 1,280 \\
BatchNorm-3 & $(128, 7, 7)$ & 256 \\
ReLU-2 & $(128, 7, 7)$ & 0 \\
Dropout-2 & $(128, 7, 7)$ & 0 \\
\midrule
Grouped Conv2D-3 ($3 \times 3$) & $(256, 5, 5)$ & 2,560 \\
BatchNorm-4 & $(256, 5, 5)$ & 512 \\
ReLU-3 & $(256, 5, 5)$ & 0 \\
Dropout-3 & $(256, 5, 5)$ & 0 \\
\midrule
Depthwise Conv2D-4 ($3 \times 3$) & $(256, 3, 3)$ & 2,560 \\
BatchNorm-5 & $(256, 3, 3)$ & 512 \\
ReLU-4 & $(256, 3, 3)$ & 0 \\
Dropout-4 & $(256, 3, 3)$ & 0 \\
\midrule
Flatten & $(2304)$ & 0 \\
\midrule
Dense-1 & $(64)$ & 147,520 \\
ReLU-5 & $(64)$ & 0 \\
\midrule
Output Layer & $(2)$ & 130 \\
\midrule
\textbf{Total Parameters} & \textbf{--} & \textbf{171,560} \\
\bottomrule
\end{tabular}
\label{tab:hybrid_cnn_architecture}
\end{table}

\subsection{Merger regression}
Table \ref{tab:cnn_architecture} shows the architecture of the CNN used as a benchmark for merger regression (Section \ref{sec:regressionidealisedresults}). Compared to the classification CNN from \cite{avirett-mackenzie_post-merger_2024}, this architecture has increased capacity to model the more complex decision boundary required for regression. It consists of five convolutional blocks with progressively increasing channel dimensions ($32$ to $512$), followed by four dense layers that reduce to a single output value. Batch normalisation and dropout are applied throughout to stabilise training and reduce overfitting. The model takes single-channel $128 \times 128$ images as input and outputs a continuous prediction for either log merger mass ratio or time since merger.

\begin{table}
\centering
\caption{Schematic table showing the layers of the CNN optimised for merger regression, as explained in Section \ref{sec:cnns}.}
\begin{tabular}{|l|c|c|}
\toprule
\textbf{Layer} & \textbf{Output Shape} & \textbf{\# Parameters} \\
\midrule
Input Layer & (1, 128, 128) & - \\
\midrule
Conv2D-1 (5 $\times$ 5) & (32, 42, 42) & 832 \\
BatchNorm-1 & (32, 42, 42) & 64 \\
ReLU-1 & (32, 42, 42) & 0 \\
Dropout-1 (0.05) & (32, 42, 42) & 0 \\
\midrule
Conv2D-2 (3 $\times$ 3) & (64, 20, 20) & 18,496 \\
BatchNorm-2 & (64, 20, 20) & 128 \\
ReLU-2 & (64, 20, 20) & 0 \\
Dropout-2 (0.05) & (64, 20, 20) & 0 \\
\midrule
Conv2D-3 (3 $\times$ 3) & (128, 18, 18) & 73,856 \\
BatchNorm-3 & (128, 18, 18) & 256 \\
ReLU-3 & (128, 18, 18) & 0 \\
MaxPool2D-3 (2 $\times$ 2) & (128, 9, 9) & 0 \\
Dropout-3 (0.1) & (128, 9, 9) & 0 \\
\midrule
Conv2D-4 (3 $\times$ 3) & (256, 7, 7) & 295,168 \\
BatchNorm-4 & (256, 7, 7) & 512 \\
ReLU-4 & (256, 7, 7) & 0 \\
Dropout-4 (0.15) & (256, 7, 7) & 0 \\
\midrule
Conv2D-5 (3 $\times$ 3) & (512, 5, 5) & 1,180,160 \\
BatchNorm-5 & (512, 5, 5) & 1,024 \\
ReLU-5 & (512, 5, 5) & 0 \\
\midrule
AdaptiveAvgPool2D & (512, 5, 5) & 0 \\
Flatten & (12800) & 0 \\
Dropout (0.3) & (12800) & 0 \\
\midrule
Dense-1 & (512) & 6,554,112 \\
BatchNorm-6 & (512) & 1,024 \\
ReLU-6 & (512) & 0 \\
Dropout-6 (0.25) & (512) & 0 \\
\midrule
Dense-2 & (256) & 131,328 \\
BatchNorm-7 & (256) & 512 \\
ReLU-7 & (256) & 0 \\
Dropout-7 (0.2) & (256) & 0 \\
\midrule
Dense-3 & (128) & 32,896 \\
BatchNorm-8 & (128) & 256 \\
ReLU-8 & (128) & 0 \\
Dropout-8 (0.2) & (128) & 0 \\
\midrule
Dense-4 & (32) & 4,128 \\
BatchNorm-9 & (32) & 64 \\
ReLU-9 & (32) & 0 \\
Dropout-9 (0.1) & (32) & 0 \\
\midrule
Output Layer & (1) & 33 \\
\midrule
\textbf{Total Parameters} & \textbf{-} & \textbf{8,294,849} \\
\bottomrule
\end{tabular}
\label{tab:cnn_architecture}
\end{table}

\section{Additional visualisations}
\label{apdx:additional-visualisations}
This appendix presents additional visualisations of the features extracted by the solid harmonic wavelet scattering bispectrum, complementing the analysis in Section \ref{sec:featureimportance}. These figures illustrate how the representation varies across different galaxy morphologies.

\subsection{Bispectral feature interactions}
\label{apdx:more-bispectral-feature-interactions}

\begin{figure}
    \centering
    \begin{subfigure}{\columnwidth}
        \centering
        \includegraphics[width=\linewidth]{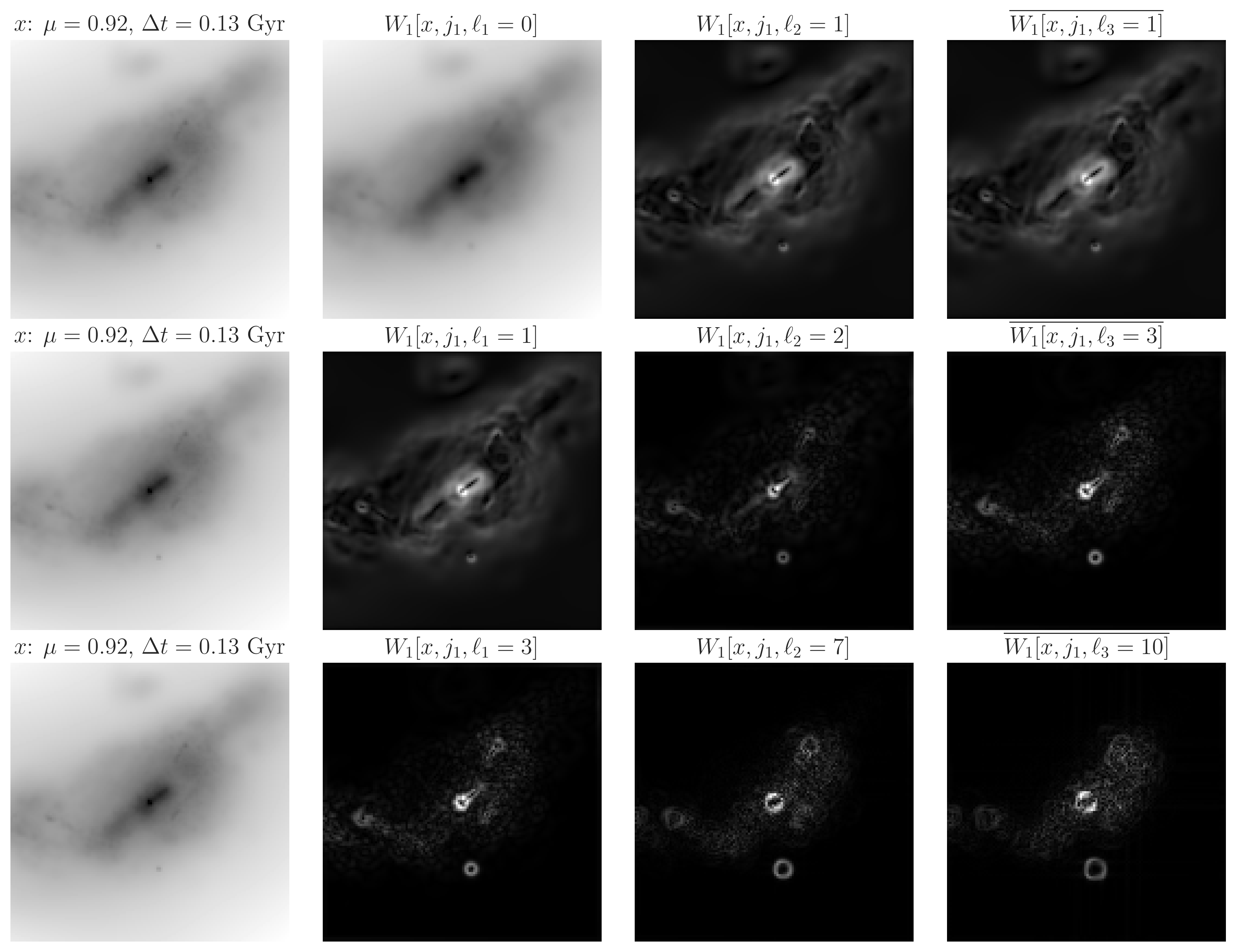}
        \caption{Example 1}
    \end{subfigure}

    \begin{subfigure}{\columnwidth}
        \centering
        \includegraphics[width=\linewidth]{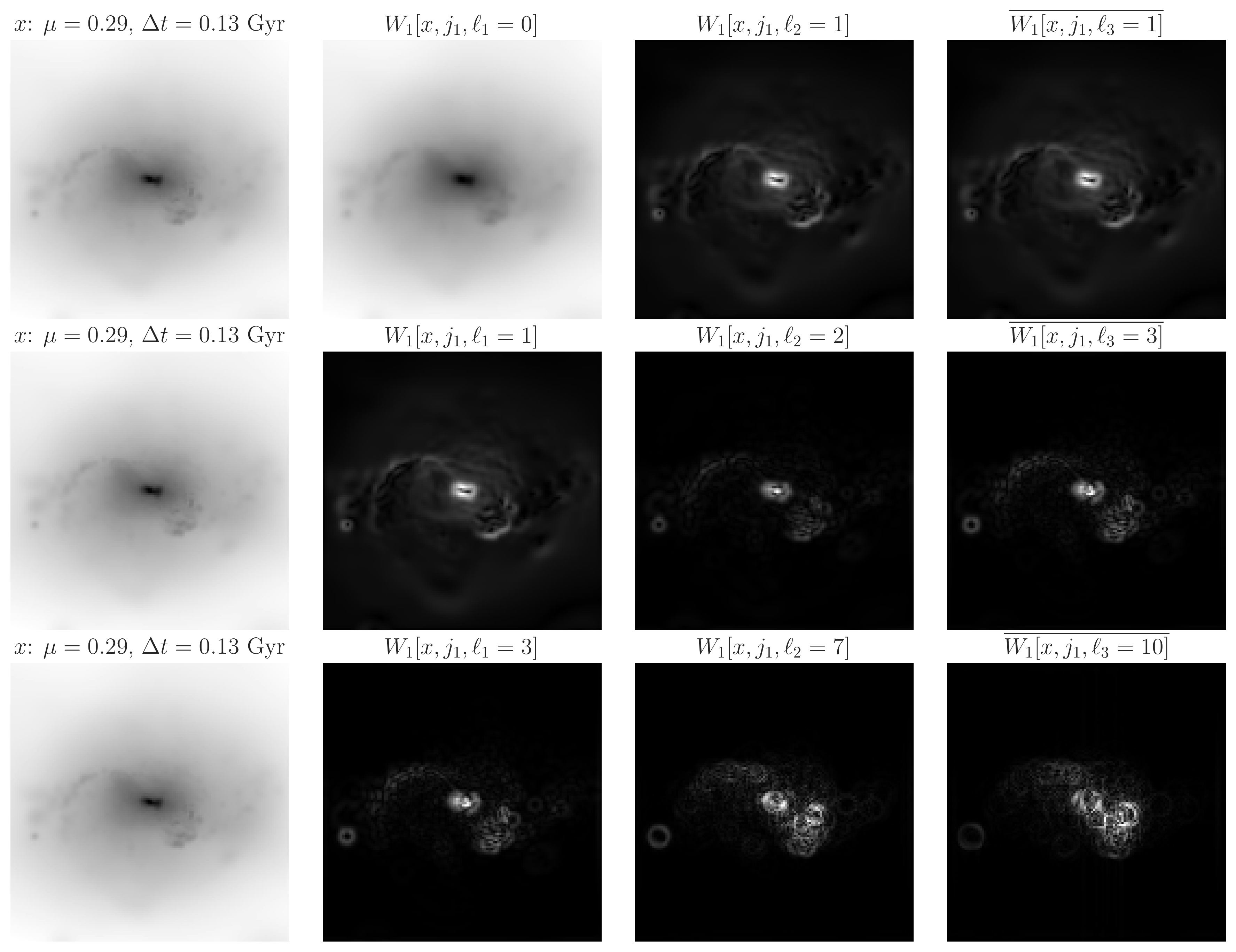}
        \caption{Example 2}
    \end{subfigure}

    \begin{subfigure}{\columnwidth}
        \centering
        \includegraphics[width=\linewidth]{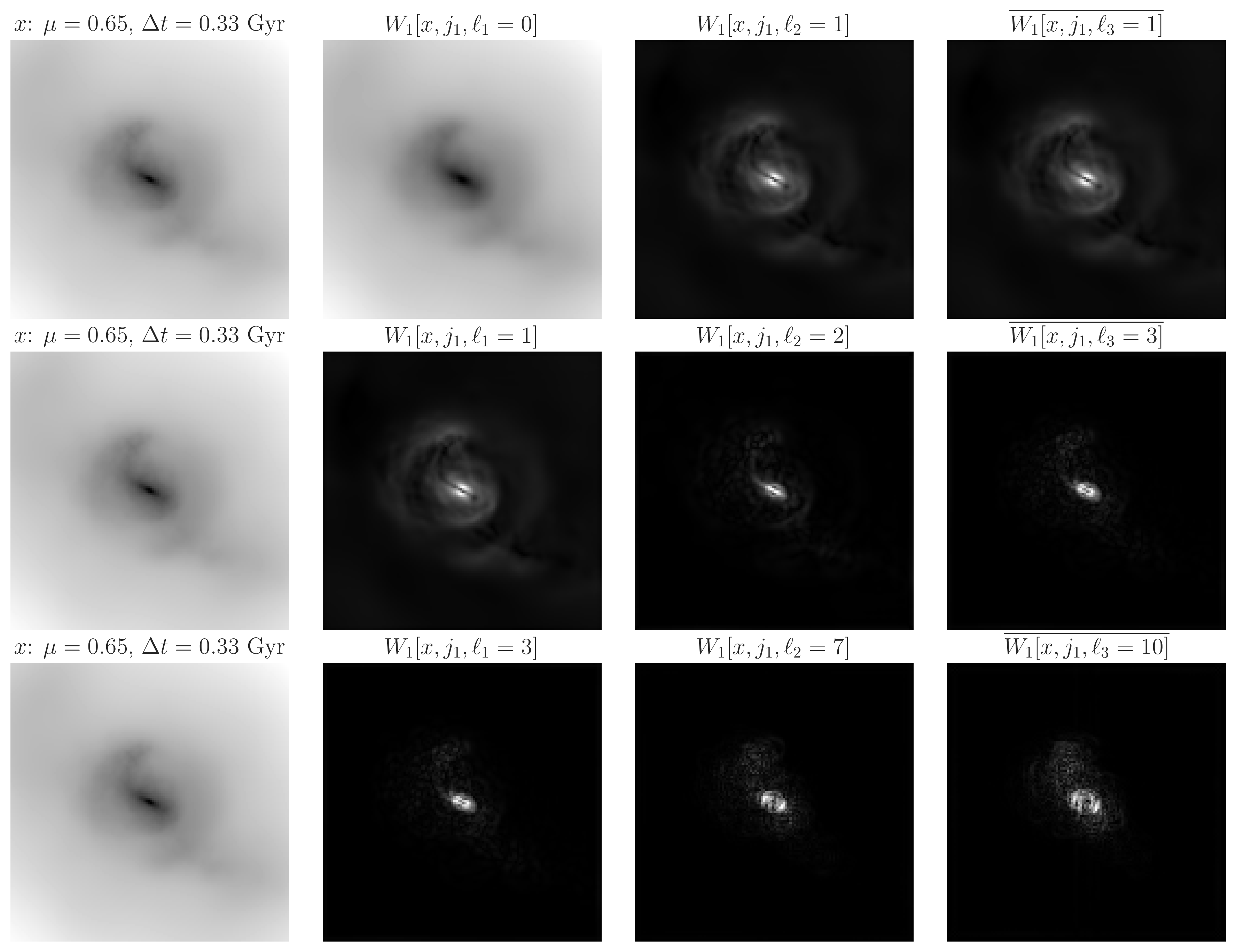}
        \caption{Example 3}
    \end{subfigure}
    
    \caption{More visualisations of features extracted for key bispectral interactions over merger samples, as in Figure \ref{fig:features-extracted-bispectral-interactions}.}
    \label{fig:more-features-extracted-bispectral-interactions}
\end{figure}

Figure \ref{fig:more-features-extracted-bispectral-interactions} extends the visualisation in Figure \ref{fig:features-extracted-bispectral-interactions} by showing the same representative $(\ell_1,\ell_2)$ interactions for three additional merger samples. The variation across galaxies demonstrates how the bispectrum captures different morphological structures depending on the underlying merger properties. For instance, the $(\ell_1=0, \ell_2=1)$  interaction consistently highlights asymmetric extended structure, while higher angular frequency combinations such as $(\ell_1=3, \ell_2=7)$ respond to finer directional patterns that vary more substantially between samples.

\begin{figure}
    \centering
    \includegraphics[width=\columnwidth]{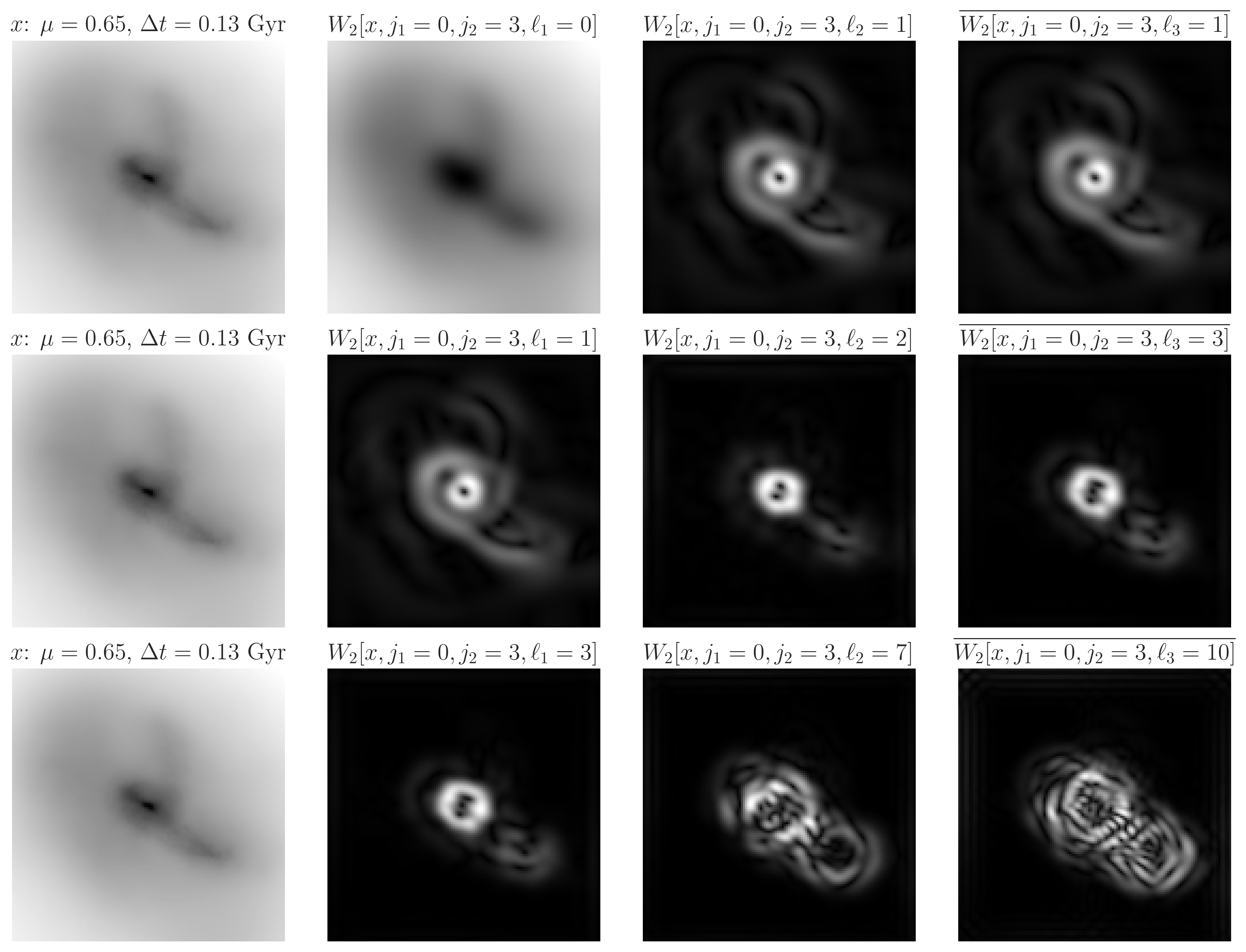}
   \caption{Visualisation of features extracted for key bispectral interactions over merger samples, as in Figure \ref{fig:features-extracted-bispectral-interactions} but for second-order scattering coefficients.}
   \label{fig:bispectral-second-order}
\end{figure}

\subsection{Scale interactions}
\label{apdx:more-second-order-features}

\begin{figure}
    \centering
    \begin{subfigure}{\columnwidth}
        \centering
        \includegraphics[width=\linewidth]{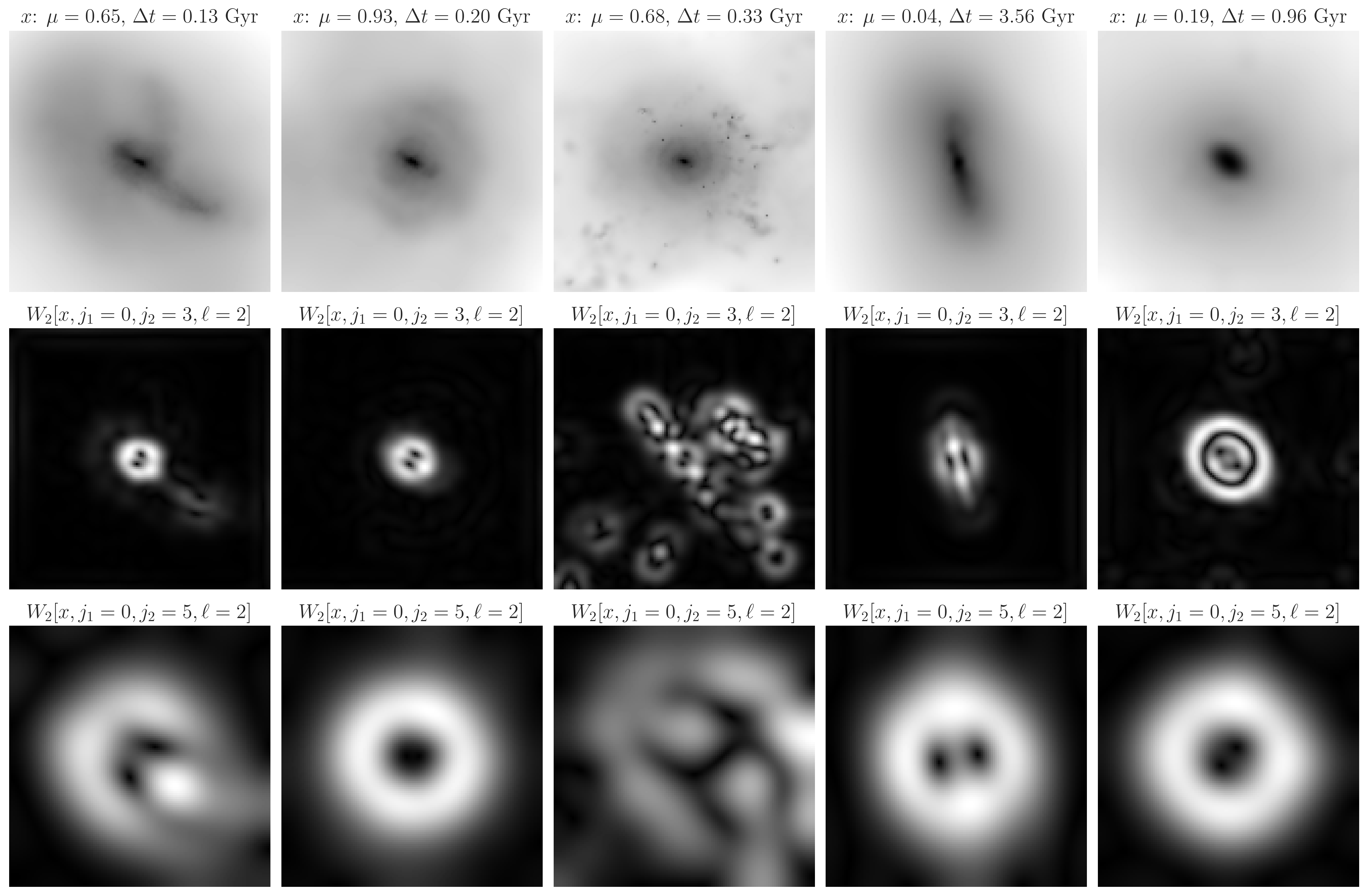}
        \caption{Example with $\ell=2$.}
    \end{subfigure}

    \begin{subfigure}{\columnwidth}
        \centering
        \includegraphics[width=\linewidth]{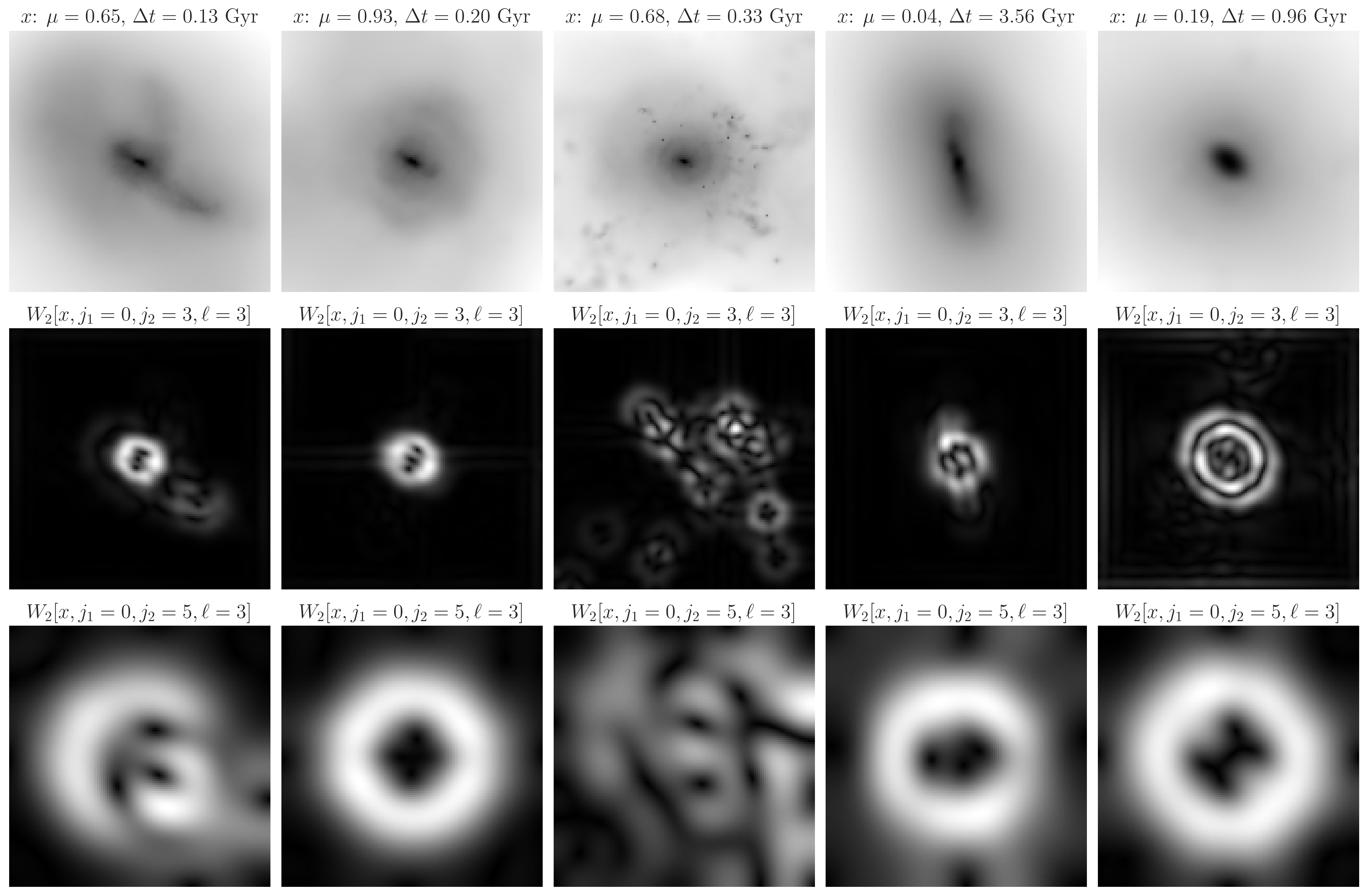}
        \caption{Example with $\ell=3$.}
    \end{subfigure}
    
    \caption{More visualisations of second-order features extracted, as in Figure \ref{fig:features-extracted-second-order}, over increasing angular frequencies.}
    \label{fig:more-features-extracted-second-order}
\end{figure}

Figure \ref{fig:more-features-extracted-second-order} extends the scale interaction visualisations from Figure \ref{fig:features-extracted-second-order} by showing second-order features at different angular frequencies ($\ell=2$ and $\ell=3$). These visualisations demonstrate that the representation captures complementary information across both scale and angular frequency, consistent with the uniform feature importance observed in Section \ref{sec:featureimportance}.

%%%%%%%%%%%%%%%%%%%%%%%%%%%%%%%%%%%%%%%%%%%%%%%%%%

\section{Independence of merger mass ratio and time since last merger}
\label{app:ratio_time}

The regression models predict the merger mass ratio $\log(\mu)$ and the time since the last merger $\Delta t$ as separate targets. Here we check that the two labels are not strongly related in our data.

We use the same samples as the regression analysis: the full sample before curation, the sample curated to be uniform in $\log(\mu)$ (used for the mass-ratio regression) and the sample curated to be uniform in $\Delta t$ (used for the time regression). Curation divides the range of the target into nine equal-width bins and randomly keeps at most 2000 samples per bin. Since the three projections of each galaxy share the same labels, every galaxy is counted once when computing the statistics below. We measure linear association with the Pearson coefficient $r$, using $r^2$ as the fraction of the variance in $\log(\mu)$ that $\Delta t$ explains linearly. We measure monotonic association with the Spearman rank coefficient $\rho$.

\begin{figure*}
    \centering
    \includegraphics[width=0.3\textwidth]{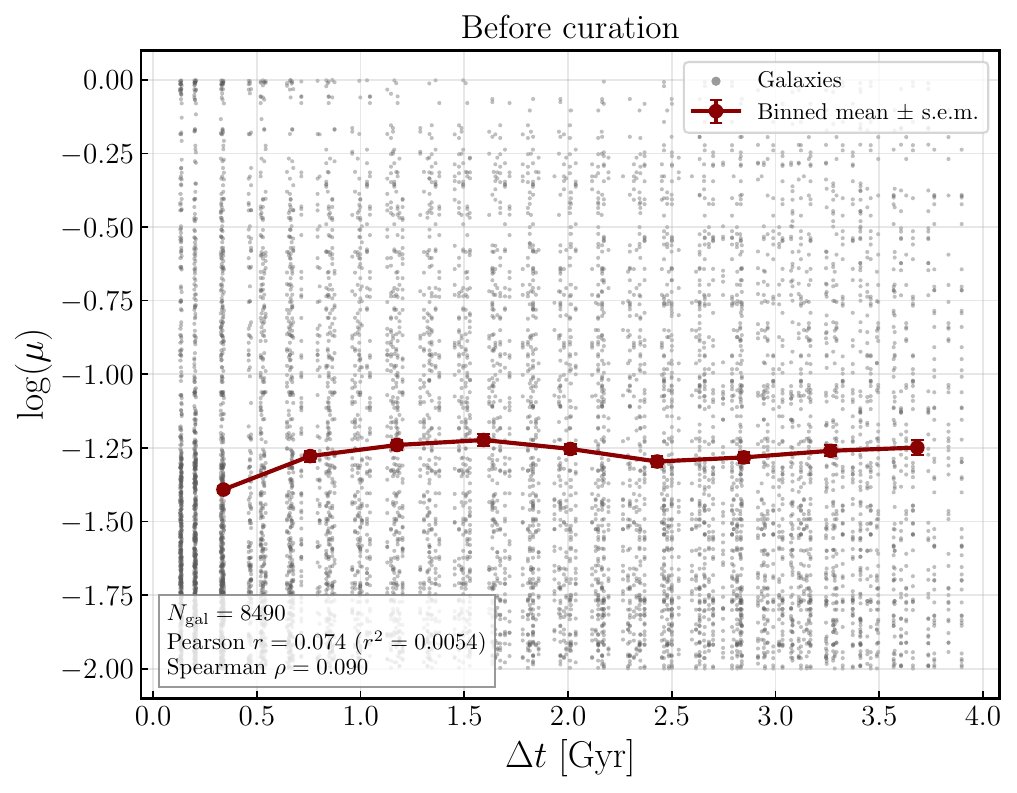}
    \includegraphics[width=0.3\textwidth]{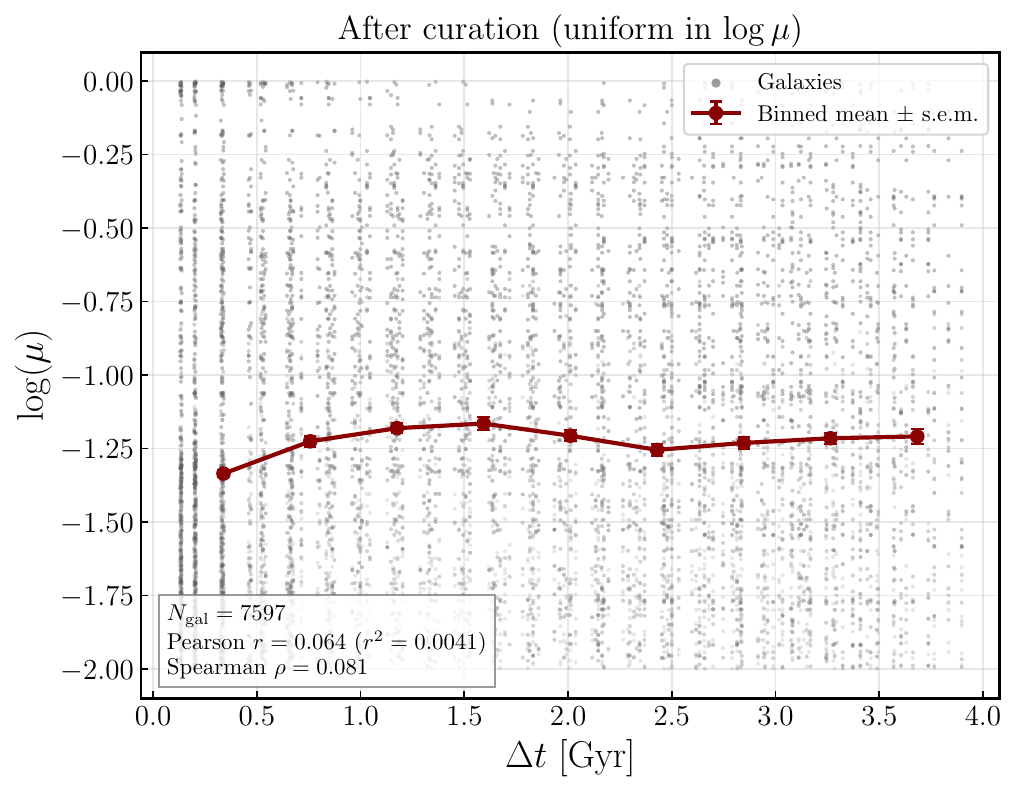}
    \includegraphics[width=0.3\textwidth]{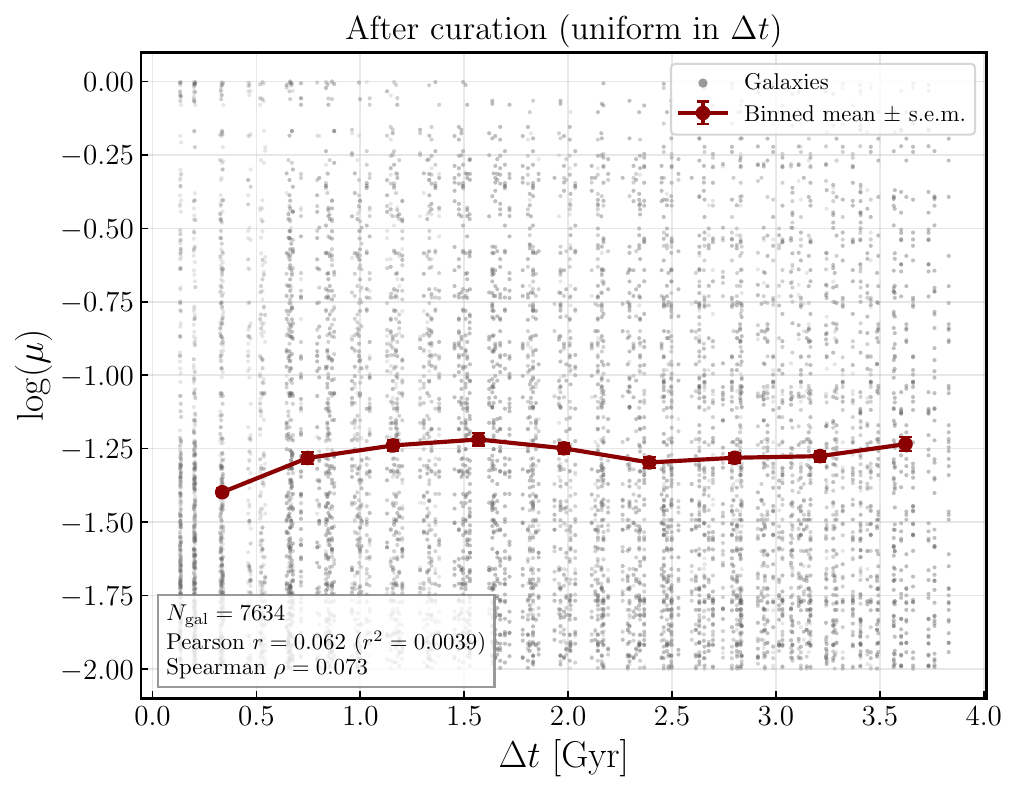}
    \caption{Merger mass ratio $\log(\mu)$ against time since last merger $\Delta t$ for the full sample (left), the sample curated to be uniform in $\log(\mu)$ (middle) and the sample curated to be uniform in $\Delta t$ (right). Grey points show individual galaxies. Red points show the mean $\log(\mu)$ in bins of $\Delta t$, with error bars giving the standard error of the mean.}
    \label{fig:ratio_time}
\end{figure*}

Figure~\ref{fig:ratio_time} shows $\log(\mu)$ against $\Delta t$ for the three samples, together with the mean $\log(\mu)$ in bins of $\Delta t$. Table~\ref{tab:ratio_time} lists the coefficients. All three samples show a very weak positive association, with $r = 0.06$--$0.07$ and $\rho = 0.07$--$0.09$. $\Delta t$ therefore accounts for at most 0.5 per cent of the variance in $\log(\mu)$. Curation does not strengthen the relation, as both coefficients are slightly smaller in the curated samples than in the full sample. We conclude that the two labels are effectively independent, so the two regression tasks can be treated as separate.

\begin{table}
    \centering
    \caption{Pearson ($r$) and Spearman ($\rho$) correlation coefficients between $\log(\mu)$ and $\Delta t$, with each galaxy counted once.}
    \label{tab:ratio_time}
    \begin{tabular}{lcccc}
        \hline
        Sample & $N_{\mathrm{gal}}$ & $r$ & $r^2$ & $\rho$ \\
        \hline
        Before curation        & 8490 & 0.074 & 0.0054 & 0.090 \\
        Uniform in $\log(\mu)$ & 7597 & 0.064 & 0.0041 & 0.081 \\
        Uniform in $\Delta t$  & 7634 & 0.062 & 0.0039 & 0.073 \\
        \hline
    \end{tabular}
\end{table}

\section{Why the predictions take the values they do}
\label{app:interpretability}
 
Saliency maps are a common way to interpret neural networks that classify or regress galaxy images. They are computed
after training and highlight the pixels the network relied on. They show where a network looks. They do not show what
it measures there, and they approximate a function whose internal features have no fixed meaning.
 
The regressor in this work takes solid-harmonic wavelet scattering coefficients as input. Each coefficient is
predefined and indexed by a wavelet scale $j$, an angular order $\ell$, a scattering order and an $L^p$ norm. With a
linear regressor, the prediction decomposes exactly into contributions of these named components. The components can
also be removed and the model retrained, which tests which of them carry the information. We use these two properties
to identify the scales that drive the predictions of $\Delta t$ and $\log\mu$ and the causes of the largest errors.
 
\subsection{Set-up}
\label{app:interp-setup}
 
All results use the 15\,s exposure images. This is the shortest exposure time before the regression performance
drops (Section~\ref{sec:regressionnoisyresults} and Fig.~\ref{fig:mae-vs-exposure-time}). We split the galaxies five times at random into training (70\%),
validation (10\%) and test (20\%) sets. All images of the same merger event stay on the same side of each split. We
quote values as mean $\pm$ standard deviation over the five test sets and test differences with the corrected
resampled $t$-test \citep{Nadeau2003}.
 
The interpretable model is a ridge regression on the standardised SHWSBic coefficients (``SHWSBic + linear''). The
coefficients are the bicoherence-normalised bispectral terms of first- and second-order scattering
(equations~\ref{SHWSBic1} and~\ref{SHWSBic2}) for $\ell \leq 10$, scales $j = 0, \dots, 6$ and $L^p$ norms with
$p = 0.5$, 1 and 2. This gives 2\,520 coefficients and 2\,521 trained parameters. Because the coefficients are
standardised, the regression weights measure the importance of each coefficient relative to the others within this
model. The penalty is chosen on the validation set. The model reaches a test MAE of $0.414 \pm 0.006$ in $\log\mu$ and
$0.842 \pm 0.023$\,Gyr in $\Delta t$.
 
For comparison, we train a convolutional neural network (CNN) under the same protocol. It follows the architecture of
\citet{avirett-mackenzie_post-merger_2024}, which was designed for $36\times36$-pixel cut-outs. Our regression images
have $128\times128$ pixels, so we insert an adaptive average-pooling layer before the dense layers. The network then
accepts the larger images with the original number of trained parameters ($1.15\times10^6$). It is smaller than the
regression CNN of Table~\ref{tab:cnn_architecture} and reaches $0.433 \pm 0.011$ and $0.871 \pm 0.030$\,Gyr. A constant
(median) prediction gives $0.467 \pm 0.004$ and $0.929 \pm 0.015$\,Gyr. The model we interpret is therefore at least as
accurate as the CNN.

\subsection{Exact decomposition of the predictions by scale}
\label{app:interp-scales}
 
For a galaxy image $x$, let $\tilde{S}_k(x) = [S_k(x) - \bar{S}_k]/\sigma_k$ be its SHWSBic coefficients standardised
with their training-set mean $\bar{S}_k$ and standard deviation $\sigma_k$. The SHWSBic + linear prediction is
\begin{equation}
  \hat{y}(x) = b + \sum_{j} c_j(x), \quad
  c_j(x) = \sum_{k \in \mathcal{K}_j} w_k\,\tilde{S}_k(x).
  \label{eq:scale-decomposition}
\end{equation}
Here $w_k$ and $b$ are the regression weights and intercept, and $\mathcal{K}_j$ contains the coefficients whose
first wavelet has scale $j$. This scale corresponds to structures of about $2^j$ pixels of $0.1$\,arcsec. That is
$2^j \times 0.11$\,kpc at the median redshift of the sample ($z = 0.056$) and between $2^j \times 0.02$ and
$2^j \times 0.19$\,kpc over $0.01 < z < 0.10$. The term $c_j(x)$ is the shift in the prediction for this galaxy,
relative to an average training galaxy, that comes from structure at scale $j$. Equation~\eqref{eq:scale-decomposition}
is an identity. It involves no approximation, no reference image and no choice of method.

\begin{figure*}
  \centering
  \includegraphics[width=\textwidth]{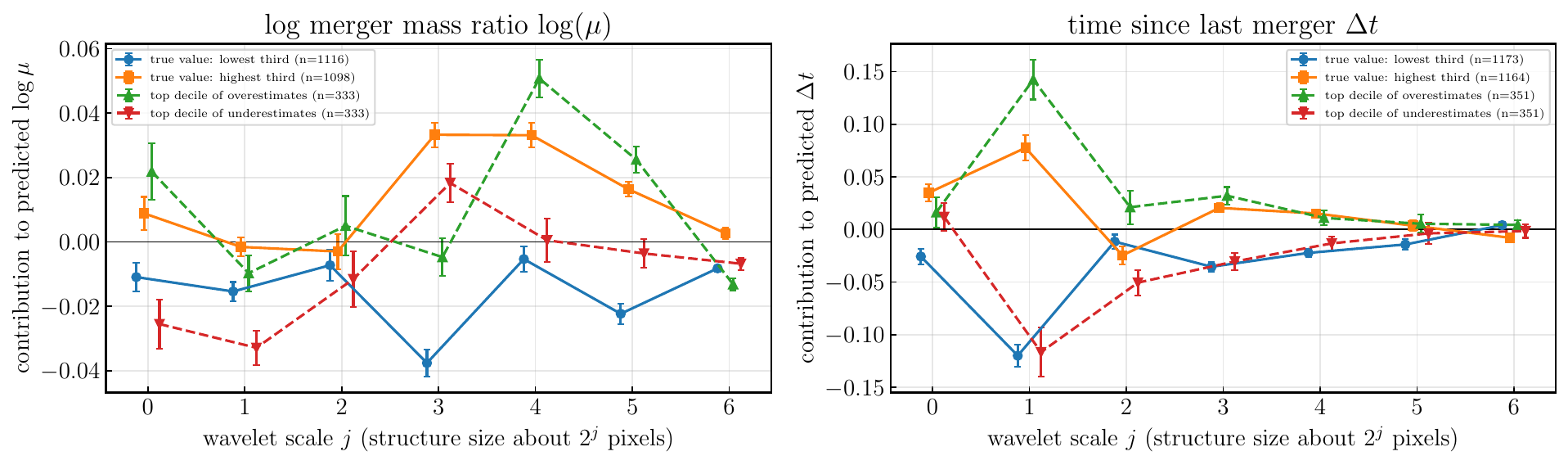}
  \caption{Exact contribution $c_j$ of each wavelet scale to the SHWS + linear prediction
  (equation~\ref{eq:scale-decomposition}), averaged over the test galaxies of one split. The groups are the lower and
  upper thirds of the true value and the 10 per cent largest over- and underestimates. Error bars are standard errors
  of the mean. Left panel $\log\mu$, right panel $\Delta t$.}
  \label{fig:scale-contributions}
\end{figure*}

Figure~\ref{fig:scale-contributions} shows $c_j$ averaged over the test set. The finest scale, $j = 1$, dominates the
prediction of $\Delta t$. Galaxies in the upper third of true $\Delta t$ receive on average $c_1 = +0.08$\,Gyr and those
in the lower third $c_1 = -0.12$\,Gyr. All other scales contribute less than $0.04$\,Gyr. At the median redshift,
$j = 1$ corresponds to $\approx 0.2$\,kpc. This is below the $0.7$\,kpc gravitational softening length of the
simulation, so these coefficients cannot resolve physical sub-kiloparsec structure. They most plausibly respond to how
sharply the light peaks at the resolution limit and to the image noise. A sharp peak means a compact, centrally
concentrated galaxy.
 
For $\log\mu$, the information sits at intermediate scales. The upper and lower thirds of true $\log\mu$ differ mainly
in $c_3$ ($+0.033$ and $-0.038$), $c_4$ ($+0.033$ and $-0.005$) and $c_5$ ($+0.016$ and $-0.022$).

\subsection{Which components carry the information}
\label{app:interp-ablation}

The representation is fixed, so we can remove a component, retrain the regressor and keep every other part of the
analysis unchanged. Table~\ref{tab:ablation} reports the results. Removing the bicoherence normalisation degrades both
targets significantly ($+0.026$ in $\log\mu$ and $+0.052$\,Gyr in $\Delta t$). This normalisation makes the
coefficients insensitive to the overall brightness of the galaxy, and it is essential. Removing the largest scale
($J = 5$) has no effect. Keeping only scales up to $J = 3$ degrades $\log\mu$ significantly. The useful information
therefore lies at scales of about $2^4$--$2^5$ pixels and below. Using a single $L^p$ norm instead of three also costs
accuracy. Finally, with a linear regressor, the 924 solid-harmonic scattering coefficients alone perform as well as
the 2\,520 bispectral coefficients. At this signal-to-noise ratio the bispectral terms add no information that a
linear model can use.
% [CHECK] This last result bears on the claims about the bispectrum in the main text.

The CNN does not allow such tests. Its intermediate features are learned, mix scales and differ between training runs.
Removing the information at a scale of $2^4$ pixels from it would require a different architecture and a new model.

\begin{table*}
  \centering
  \small
  \caption{Test MAE of SHWS + linear when parts of the representation are removed and the regressor is retrained.
  Values are mean $\pm$ s.d.\ over five grouped splits. $\Delta$ is the change with respect to the reference and $p$
  the corrected resampled $t$-test. $J$ is the largest wavelet scale kept.}
  \label{tab:ablation}
  \begin{tabular}{lrcccc}
    \toprule
    & & \multicolumn{2}{c}{$\log\mu$} & \multicolumn{2}{c}{$\Delta t$ [Gyr]} \\
    Representation & Coeff. & MAE & $\Delta$ ($p$) & MAE & $\Delta$ ($p$) \\
    \midrule
    Reference (bispectrum, bicoherence) & 2\,520 & $0.414 \pm 0.006$ & -- & $0.842 \pm 0.023$ & -- \\
    No bicoherence normalisation & 2\,520 & $0.440 \pm 0.010$ & $+0.026$ (0.008) & $0.894 \pm 0.021$ & $+0.052$ ($<0.001$) \\
    Scattering coefficients only & 924 & $0.416 \pm 0.006$ & $+0.002$ (0.51) & $0.843 \pm 0.027$ & $+0.001$ (0.72) \\
    $J = 5$ & 1\,890 & $0.414 \pm 0.004$ & $-0.000$ (0.89) & $0.844 \pm 0.025$ & $+0.002$ (0.24) \\
    $J = 4$ & 1\,350 & $0.416 \pm 0.007$ & $+0.002$ (0.14) & $0.845 \pm 0.026$ & $+0.003$ (0.53) \\
    $J = 3$ & 900 & $0.420 \pm 0.007$ & $+0.007$ (0.010) & $0.855 \pm 0.027$ & $+0.013$ (0.087) \\
    Single $L^1$ norm & 840 & $0.419 \pm 0.006$ & $+0.006$ (0.004) & $0.848 \pm 0.025$ & $+0.006$ (0.052) \\
    \bottomrule
  \end{tabular}
\end{table*}

\subsection{Interpretation of specific cases with high errors}
\label{app:interp-failures}

The scale decomposition also explains the failures. For $\Delta t$, the 10 per cent largest overestimates receive on
average $c_1 = +0.14$\,Gyr. Galaxies that truly merged long ago receive less, $+0.08$\,Gyr for the upper third of true
$\Delta t$. The 10 per cent largest underestimates receive $c_1 = -0.12$\,Gyr, the same as galaxies that truly merged
recently (lower third, Fig.~\ref{fig:scale-contributions}). The cases with high discrepancies are therefore galaxies whose finest-scale
structure resembles that of galaxies at the opposite end of the $\Delta t$ range.

\begin{figure*}
  \centering
  \includegraphics[width=\textwidth]{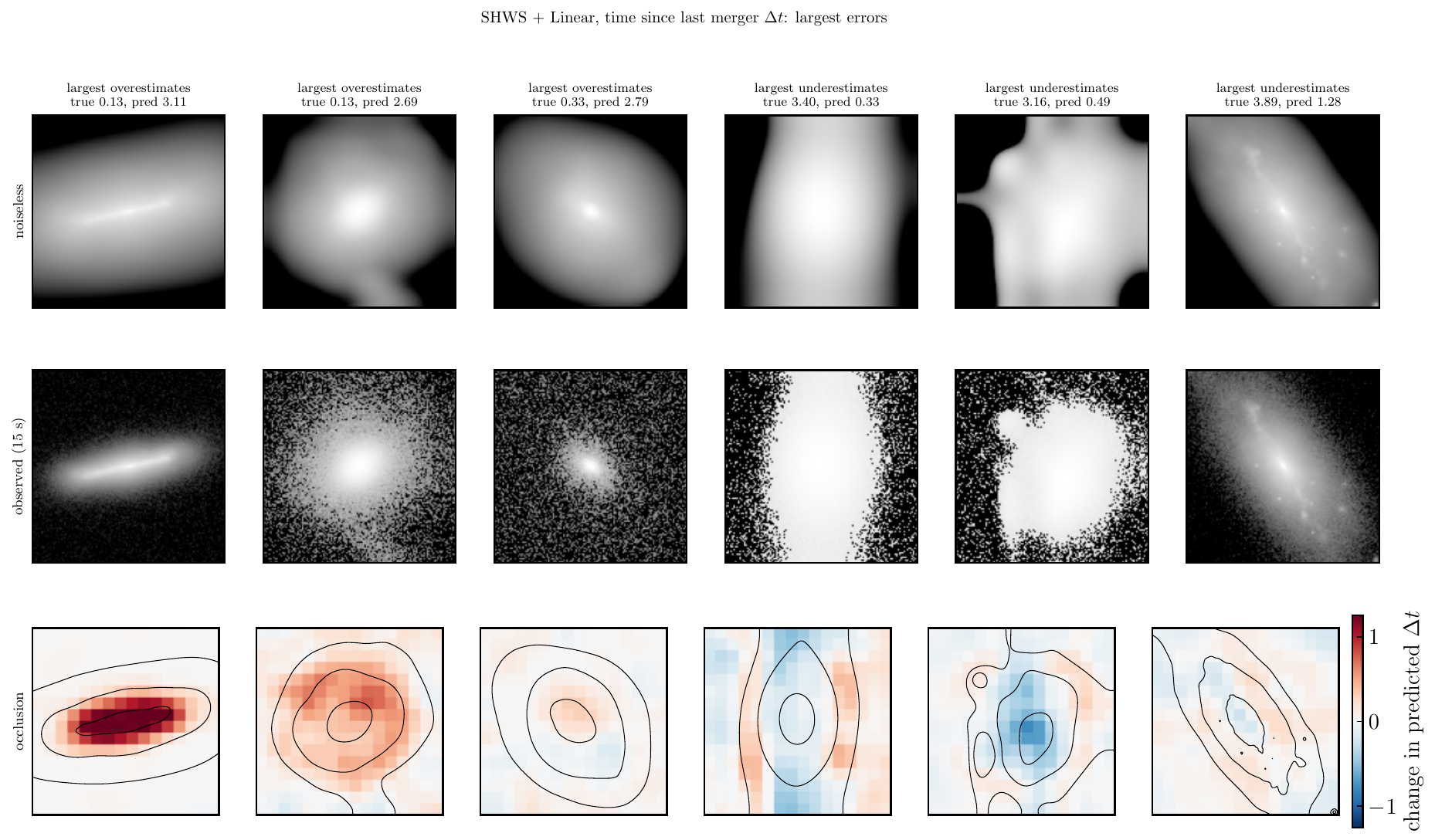}
  \caption{The three largest overestimates (left) and underestimates (right) of $\Delta t$, each from a different
  merger event. The rows show the noiseless image, the observed 15\,s image and the occlusion map of the SHWS + linear
  prediction with the noiseless isophotes overlaid. Red regions raise the predicted $\Delta t$ and blue regions
  lower it.}
  \label{fig:failures}
\end{figure*}

\begin{figure*}
  \centering
  \includegraphics[width=\textwidth]{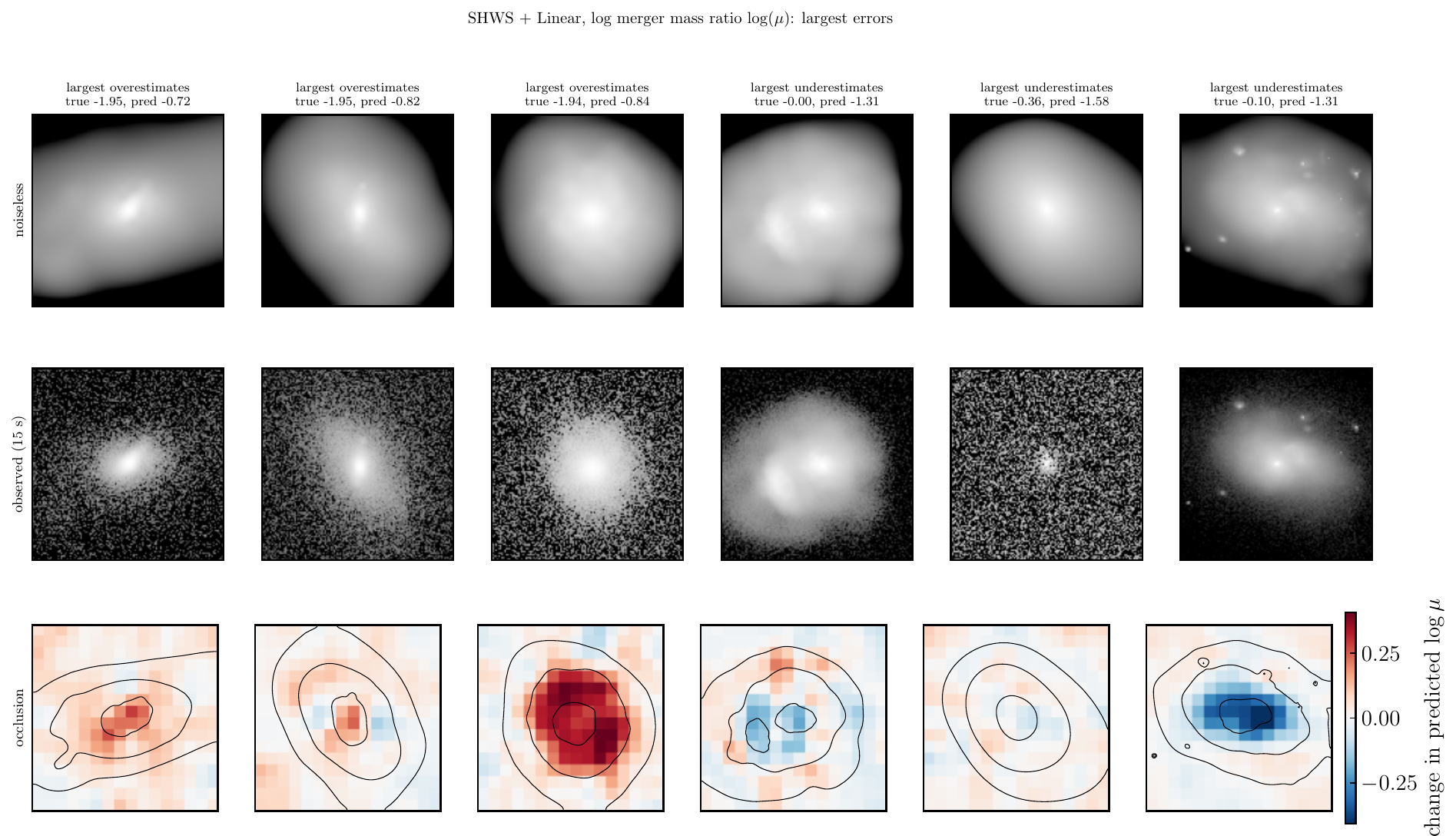}
  \caption{As Fig.~\ref{fig:failures}, for $\log\mu$.}
  \label{fig:failures-ratio}
\end{figure*}

Figures~\ref{fig:failures} and~\ref{fig:failures-ratio} show the largest failures, each from a different merger event.
Each comes with an occlusion map of the SHWS + linear prediction \citep{Zeiler2014}. To compute it, we replace each
$16\times16$-pixel patch with sky pixels drawn from the image border and record how much the prediction drops. Red
regions raise the predicted value and blue regions lower it.

The three largest overestimates of $\Delta t$ are recent mergers ($\Delta t = 0.13$--$0.33$\,Gyr) predicted at
2.7--3.1\,Gyr. One is a thin edge-on disc, and its occlusion map is strongly red along the disc. The other two are
symmetric galaxies, only one shows asymmetric extended features.
The three largest underestimates are old mergers ($\Delta t = 3.2$--$3.9$\,Gyr) predicted at 0.3--1.3\,Gyr. The show asymmetric structures, and one is a disc with clumpy star-forming regions, all potential features found in recent mergers.

For $\log\mu$, intermediate scales drive the overestimates of the mass ratio. They receive $c_4 = +0.05$, compared with $+0.03$ for
galaxies in the upper third of true $\log\mu$. The finest scales drive the underestimates, with $c_0 = -0.03$ and
$c_1 = -0.03$ compared with $-0.01$ and $-0.02$ for the lower third. In the examples (Fig.~\ref{fig:failures-ratio}),
the largest overestimates are minor mergers ($\log\mu \approx -1.95$) predicted at $-0.7$ to $-0.8$. They are relatively compact, with some extended asymmetries. At high mass ratios, underestimates occur, this is expected given the physical upper limit on the mass ratio. Of the underestimates in major mergers, one shows clear asymmetries. One is a low signal to noise image. One is a disc with clumps.

\subsection{Limitations}
\label{app:interp-limitations}

We chose the linear regressor because its predictions
decompose exactly, not because it gives the highest accuracy. Non-linear regressors on the same coefficients, such as
the KNN regressor of Section~\ref{sec:regressionidealisedresults}, may weight the coefficients differently, and the
decomposition of equation~\eqref{eq:scale-decomposition} does not apply to them.
 
The occlusion maps have the resolution of the 16-pixel occluding patch. They locate discs but cannot resolve
individual spiral arms.

% Don't change these lines
\bsp	% typesetting comment
\label{lastpage}
\end{document}